\documentclass[graybox,multphys,norunningheads]{svmult}

\usepackage{titlesec}
\usepackage{soul}
\usepackage{geometry}
\usepackage{type1cm}       
\usepackage{cite}
\usepackage{makeidx}         
\usepackage{graphicx}       
                             
\usepackage{multicol}        
\usepackage[bottom]{footmisc}

\usepackage{newtxtext}       
\usepackage{newtxmath}       

\usepackage[pdfpagemode={UseOutlines},bookmarks=true,bookmarksopen=true,
bookmarksopenlevel=0,bookmarksnumbered=true,hypertexnames=false,
colorlinks,linkcolor={blue},citecolor={blue},urlcolor={black},
pdfstartview={FitV},unicode,breaklinks=true,pagebackref,hyperindex=true]{hyperref}
\usepackage{booktabs}
\usepackage{array}
\graphicspath{{Figures/}}
\newcolumntype{C}[1]{>{\centering\arraybackslash$}p{#1}<{$}}

\usepackage[utf8]{inputenc}
\usepackage[english]{babel}
\usepackage[T1]{fontenc}

\usepackage{xcolor}
\definecolor{maroon}{cmyk}{0, 0.87, 0.68, 0.32}
\definecolor{halfgray}{gray}{0.55}
\definecolor{ipython_frame}{RGB}{207, 207, 207}
\definecolor{ipython_bg}{RGB}{247, 247, 247}
\definecolor{ipython_red}{RGB}{186, 33, 33}
\definecolor{ipython_green}{RGB}{0, 128, 0}
\definecolor{ipython_cyan}{RGB}{64, 128, 128}
\definecolor{ipython_purple}{RGB}{170, 34, 255}

\usepackage{listings}
\lstdefinelanguage{iPython}{
	morekeywords={access,and,break,class,continue,def,del,elif,else,except,exec,finally,for,from,global,if,import,in,is,lambda,not,or,pass,print,raise,return,try,while},%
	morekeywords=[2]{abs,all,any,basestring,bin,bool,bytearray,callable,chr,classmethod,cmp,compile,complex,delattr,dict,dir,divmod,enumerate,eval,execfile,file,filter,float,format,frozenset,getattr,globals,hasattr,hash,help,hex,id,input,int,isinstance,issubclass,iter,len,list,locals,long,map,max,memoryview,min,next,object,oct,open,ord,pow,property,range,raw_input,reduce,reload,repr,reversed,round,set,setattr,slice,sorted,staticmethod,str,sum,super,tuple,type,unichr,unicode,vars,xrange,zip,apply,buffer,coerce,intern},%
	sensitive=true,%
	morecomment=[l]\#,%
	morestring=[b]',%
	morestring=[b]",%
	morestring=[s]{'''}{'''},%
	morestring=[s]{"""}{"""},%
	morestring=[s]{r'}{'},%
	morestring=[s]{r"}{"},%
	morestring=[s]{r'''}{'''},%
	morestring=[s]{r"""}{"""},%
	morestring=[s]{u'}{'},%
	morestring=[s]{u"}{"},%
	morestring=[s]{u'''}{'''},%
	morestring=[s]{u"""}{"""},%
	literate=
	{á}{{\'a}}1 {é}{{\'e}}1 {í}{{\'i}}1 {ó}{{\'o}}1 {ú}{{\'u}}1
	{Á}{{\'A}}1 {É}{{\'E}}1 {Í}{{\'I}}1 {Ó}{{\'O}}1 {Ú}{{\'U}}1
	{à}{{\`a}}1 {è}{{\`e}}1 {ì}{{\`i}}1 {ò}{{\`o}}1 {ù}{{\`u}}1
	{À}{{\`A}}1 {È}{{\'E}}1 {Ì}{{\`I}}1 {Ò}{{\`O}}1 {Ù}{{\`U}}1
	{ä}{{\"a}}1 {ë}{{\"e}}1 {ï}{{\"i}}1 {ö}{{\"o}}1 {ü}{{\"u}}1
	{Ä}{{\"A}}1 {Ë}{{\"E}}1 {Ï}{{\"I}}1 {Ö}{{\"O}}1 {Ü}{{\"U}}1
	{â}{{\^a}}1 {ê}{{\^e}}1 {î}{{\^i}}1 {ô}{{\^o}}1 {û}{{\^u}}1
	{Â}{{\^A}}1 {Ê}{{\^E}}1 {Î}{{\^I}}1 {Ô}{{\^O}}1 {Û}{{\^U}}1
	{œ}{{\oe}}1 {Œ}{{\OE}}1 {æ}{{\ae}}1 {Æ}{{\AE}}1 {ß}{{\ss}}1
	{ç}{{\c c}}1 {Ç}{{\c C}}1 {ø}{{\o}}1 {å}{{\r a}}1 {Å}{{\r A}}1
	{€}{{\EUR}}1 {£}{{\pounds}}1,
	literate=
	*{+}{{{\color{ipython_purple}+}}}1
	{-}{{{\color{ipython_purple}-}}}1
	{*}{{{\color{ipython_purple}*}}}1
	{/}{{{\color{ipython_purple}/}}}1
	{^}{{{\color{ipython_purple}\^{}}}}1
	{?}{{{\color{ipython_purple}?}}}1
	{!}{{{\color{ipython_purple}!}}}1
	{\%}{{{\color{ipython_purple}\%}}}1
	{<}{{{\color{ipython_purple}<}}}1
	{>}{{{\color{ipython_purple}>}}}1
	{|}{{{\color{ipython_purple}|}}}1
	{\&}{{{\color{ipython_purple}\&}}}1
	{~}{{{\color{ipython_purple}~}}}1
	{==}{{{\color{ipython_purple}==}}}2
	{<=}{{{\color{ipython_purple}<=}}}2
	{>=}{{{\color{ipython_purple}>=}}}2
	{+=}{{{+=}}}2
	{-=}{{{-=}}}2
	{*=}{{{$^\ast$=}}}2
	{/=}{{{/=}}}2,
	commentstyle=\color{ipython_cyan}\ttfamily,
	stringstyle=\color{ipython_red}\ttfamily,
	keepspaces=true,
	showspaces=false,
	showstringspaces=false,
	rulecolor=\color{ipython_frame},
	frame=single,
	frameround={t}{t}{t}{t},
	framexleftmargin=6mm,
	numbers=left,
	numberstyle=\tiny\color{halfgray},
	backgroundcolor=\color{ipython_bg},
	basicstyle=\scriptsize\ttfamily,
	keywordstyle=\color{ipython_green}\ttfamily,
	escapechar=\¢,escapebegin=\color{ipython_green},
}

\makeindex             

\begin{document}

\title*{The Non-Equilibrium Green Function (NEGF) Method}
\author{Kerem Y. Camsari, Shuvro Chowdhury \\ and Supriyo Datta  }
\institute{Kerem Y. Camsari  \at Department of Electrical and Computer Engineering, University of California, Santa Barbara, CA, 93106, USA, \email{camsari@ucsb.edu}
\and Shuvro Chowdhury, Supriyo Datta \at School of Electrical and Computer Engineering, Purdue University, West Lafayette, IN 47907, USA, \email{chowdhu7@purdue.edu,datta@purdue.edu}}

\maketitle

\abstract*{The Non-Equilibrium Green Function (NEGF) method was established in the 1960's through the classic work of Schwinger, Kadanoff, Baym, Keldysh and others using many-body perturbation theory (MBPT) and the diagrammatic theory for non-equilibrium processes. Much  of the literature is based on the original MBPT-based approach and this makes it inaccessible to those unfamiliar with advanced quantum statistical mechanics. We obtain the NEGF equations directly from a one-electron Schr\"{o}dinger equation using relatively elementary arguments. These equations have been used to discuss many problems of great interest such as quantized conductance, (integer) quantum Hall effect, Anderson localization, resonant tunneling and spin transport without a systematic treatment of many-body effects. But it goes beyond purely coherent transport allowing us to include phase-breaking interactions (both momentum-relaxing and momentum-conserving as well as spin-conserving and spin-relaxing) within a self-consistent Born approximation. We believe that the scope and utility of the NEGF equations  transcend the MBPT-based approach originally used to derive it. NEGF teaches us how to combine quantum dynamics with ``contacts'' much as Boltzmann taught us how to combine classical dynamics with ``contacts'', using the word ``contacts'' in a broad figurative sense to denote all kinds of entropy-driven processes. We believe that this approach to ``contact-ing'' the Schr\"{o}dinger  equation should be of broad interest to anyone working on device physics or non-equilibrium statistical mechanics in general.}

\abstract{The Non-Equilibrium Green Function (NEGF) method was established in the 1960's through the classic work of Schwinger, Kadanoff, Baym, Keldysh and others using many-body perturbation theory (MBPT) and the diagrammatic theory for non-equilibrium processes. Much  of the literature is based on the original MBPT-based approach and this makes it inaccessible to those unfamiliar with advanced quantum statistical mechanics. We obtain the NEGF equations directly from a one-electron Schr\"{o}dinger equation using relatively elementary arguments. These equations have been used to discuss many problems of great interest such as quantized conductance, (integer) quantum Hall effect, Anderson localization, resonant tunneling and spin transport without a systematic treatment of many-body effects. But it goes beyond purely coherent transport allowing us to include phase-breaking interactions (both momentum-relaxing and momentum-conserving as well as spin-conserving and spin-relaxing) within a self-consistent Born approximation. We believe that the scope and utility of the NEGF equations  transcend the MBPT-based approach originally used to derive it. NEGF teaches us how to combine quantum dynamics with ``contacts'' much as Boltzmann taught us how to combine classical dynamics with ``contacts'', using the word ``contacts'' in a broad figurative sense to denote all kinds of entropy-driven processes.  We believe that this approach to ``contact-ing'' the Schr\"{o}dinger  equation should be of broad interest to anyone working on device physics or non-equilibrium statistical mechanics in general.}

\begingroup
\let\cleardoublepage\clearpage
\tableofcontents
\endgroup

\clearpage
\newpage

\section{Introduction}

The non-equilibrium Green's function\index{non-equilibrium Green's function} (NEGF)\index{NEGF} method was pioneered in the 1960's by the classic work of Martin, Schwinger \cite{martin1959theory}, Kadanoff, Baym \cite{kadanoff1989quantum}, Keldysh \cite{keldysh1965diagram} and others, which have been discussed in many review articles such as Danielewicz \cite{danielewicz1984quantum} and Mahan \cite{mahan1987quantum}. After the advent of mesoscopic physics in the 1980's this method was combined with the Landauer approach (See Refs. \cite{datta1989steady}, \cite{mclennan1991},  \cite{meir1992landauer}) and this ``NEGF-Landauer method'' has been widely used in the field of nanoelectronics for device modeling and technology development (See 
\cite{klimeck1995quantum,bowen1997quantitative,lake1997single,martinez2007self,datta2002non,seoane2009current,wang2004three,PhysRevB.68.245406,nikolic2006imaging,kubis2011assessment,wang2006nonequilibrium,koswatta2007nonequilibrium,pourfath2014non,guo2004atomistic,martinez20093}  as a small subset of a vast literature). The chapter by Klimeck and Boykin in this volume provides an introduction to state-of-the-art quantum transport simulation tools based on this method. The present chapter is intended to serve a complementary purpose, namely to introduce the conceptual underpinnings of NEGF and to answer some frequently asked questions regarding them.

\begin{figure}[!ht]
	\sidecaption
	\includegraphics[]{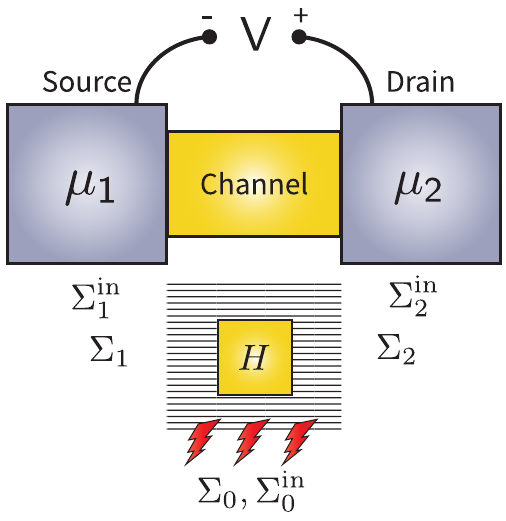}
	\caption{\textbf{A resistor:} Physical structure comprising a channel with two contacts, labeled source and drain. In the non-equilibrium Green function (NEGF) method the channel is described by a Hamiltonian $\mathbf{H}$ while its connection to the ``contacts'' are described by the self-energy functions $\pmb{\mathbf{\Sigma}}$, $\pmb{\mathbf{\Sigma}}^{\text{in}}$: subscripts 1 and 2 denote the physical contacts labeled source and drain, while the subscript 0 represents abstract contacts like the phonon bath.}
	\label{fig:figure1}
\end{figure}

The iconic device in modern electronics is the field effect transistor (FET)\index{field effect transistor (FET)}, billions of which are a part of every smartphone. An oversimplified sketch of an FET is shown in Fig.~\ref{fig:figure1} consisting of an active region marked the \emph{channel} sandwiched between two highly conductive regions labeled the \emph{source}\index{source} and the \emph{drain}\index{drain}.  An FET also has a third terminal (not shown) which can be used to control the resistance\index{resistance} $R=V/I$ of the channel\index{channel}, but we will not get into such ``details''.  We will focus just on the problem of calculating the steady-state charge current $I$ for a given voltage $V$ taking \emph{quantum} effects into account.

Quantum mechanical descriptions usually start from a Hamiltonian\index{Hamiltonian} matrix $\mathbf{H}$ whose eigenvalues\index{eigenvalues} give us the allowed energy levels. However, if we treat the structure like a closed system described just by an  $\mathbf{H}$, an applied voltage will cause opposite charges to pile up in the source and drain, and we will have a capacitor and not a resistor. What makes it a resistor is the external battery that continually takes electrons out of the drain and inserts electrons back into the source to maintain an electrochemical potential difference\index{electrochemical potential difference}: $\mu_1-\mu_2=qV$.

One way to describe such an open system is to describe the two contacts denoted 1 and 2 through self-energy functions\index{self-energy function} $\pmb{\mathbf{\Sigma}}_{1,2}$, $\pmb{\mathbf{\Sigma}}^{\text{in}}_{1,2}$. The first of these $\pmb{\mathbf{\Sigma}}_{1,2}$ has a Hermitian part that modifies the Hamiltonian and an anti-Hermitian part that represents the rate at which electrons escape from the channel. The second of these $\pmb{\mathbf{\Sigma}}^{\text{in}}_{1,2}$ represents the inflow of the electrons into the channel from the contacts. In addition to these physical contacts\index{contact} there are abstract ``contacts''  representing the interactions of the electrons with the lattice and with other electrons as they traverse the channel and these are described the self-energy functions\index{self-energy function} $\pmb{\mathbf{\Sigma}}_{0}$, $\pmb{\mathbf{\Sigma}}^{\text{in}}_{0}$. To apply the NEGF\index{NEGF} method to a given problem, there are two steps:
\begin{enumerate}
\item Identify the appropriate Hamiltonian\index{Hamiltonian} $\mathbf{H}$ and self-energy functions\index{self-energy function} $\pmb{\mathbf{\Sigma}}_{m}$, $\pmb{\mathbf{\Sigma}}_{m}^{\text{in}}$
\item Use these in the NEGF Eqs.~(\ref{eq:eq1}-\ref{eq:eq3}) summarized in Sect.~\ref{sec:negf_summary} to calculate quantities of interest like current\index{current}, electron density\index{electron density} or density of states\index{density of states}.
\end{enumerate}

The NEGF\index{NEGF} method is commonly viewed as an esoteric tool accessible only to a small group of specialists. To quote from two popular books \cite{di2008electrical,heikkila2013physics}: 
\begin{quotation}
 \ldots In this respect, the most difficult topic is probably the non-equilibrium Green's function formalism of Chapter 4 \ldots
\end{quotation}
\begin{quotation}
	 \ldots Because of my aim of avoiding too heavy formalism I have chosen not to describe non-equilibrium Green's function approaches to transport phenomena. Courses detailing these approaches usually spend half the time on the formalism, finding the poles of the various Green's functions, and figuring out analytic continuations and so on \ldots
\end{quotation}

Both books are pedagogically outstanding  and we cite them only to stress that NEGF is considered difficult even by the foremost practitioners in the field. And the reason is that the traditional discussion of NEGF is based on many-body perturbation theory\index{many-body perturbation theory} \index{MBPT}(MBPT) which takes many semesters of advanced quantum mechanics to master.

However, as we have often noted \cite{datta2015non}, the only aspect of NEGF that really requires MBPT\index{MBPT} is in writing down the self-energy functions\index{self-energy function} $\pmb{\mathbf{\Sigma}}_{0},\pmb{\mathbf{\Sigma}}^{\text{in}}_{0}$ describing interactions within the channel. Everything else can be accomplished without the use of MBPT, including deriving the NEGF Eqs.~(\ref{eq:eq1}-\ref{eq:eq3}) and writing down the non-interacting channel Hamiltonian\index{non-interacting channel Hamiltonian} $\mathbf{H}$ along with the self-energy functions for the physical contacts $\pmb{\mathbf{\Sigma}}_{1,2}$ and $\pmb{\mathbf{\Sigma}}_{1,2}^{\text{in}}$. This is the approach one of us has used in all his books starting from  Chapter 8 of Ref.~\cite{datta1997electronic} and later in  Ref.~\cite{datta2005quantum} and Part B of Ref.~\cite{datta2012lessons}. The recent book by Ghosh (Ref.~\cite{ghosh2016nanoelectronics}) is unique in describing both our approach and the MBPT-based one along with special non-perturbative approaches. 

\subsection{Decoupling NEGF from MBPT}

Why is the standard treatment so heavily dependent on MBPT? We believe the reason is historical. Until 1990 it was common to regard the physical contacts\index{contact} as an unimportant detail and view the channel interactions as the essential physics. And so the original papers in the field dating back to the 1960's described the resistance in terms of $\pmb{\mathbf{\Sigma}}_{0}$, $\pmb{\mathbf{\Sigma}}^{\text{in}}_{0}$ and never even considered the $\pmb{\mathbf{\Sigma}}_{1,2}$, $\pmb{\mathbf{\Sigma}}^{\text{in}}_{1,2}$ describing the physical contacts.

What changed around 1990 was the widespread experimental measurement of the resistance of \index{ballistic conductor}\emph{ballistic conductors} where electrons zip through the channel like bullets with little or no interactions making $\pmb{\mathbf{\Sigma}}_{0}$ and $\pmb{\mathbf{\Sigma}}^{\text{in}}_{0}$ negligible. Do such conductors have a resistance\index{resistance}? The experimental answer was a resounding \emph{yes}\begin{footnote}{For example Ref. \cite{van1988quantized} is one of the first experiments that established the fundamental limits of conductance in such ballistic conductors.  Later, scientists have even measured the resistance of a single molecule \cite{xu2003measurement}!}\end{footnote}. A current $I$ flowing through a non-zero resistance $R$ must generate a heat $I^2R$. Heat is generated when energetic electrons give up some of their energy to the lattice making the atoms jiggle, which manifests itself as an increase in temperature.

\begin{figure}[!ht]
	\centering
	\includegraphics[width=\linewidth,keepaspectratio]{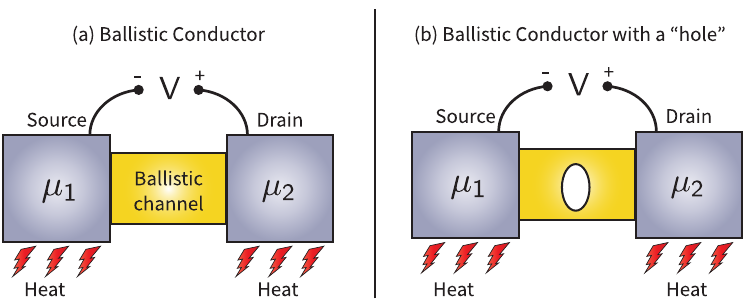}
	\caption{\textbf{Left}: In a ballistic conductor the resistance arises from the channel-contact interfaces but the heating occurs deeper in the contacts. \textbf{Right}: A ballistic conductor\index{ballistic conductor} with a hole further accentuates the spatial separation between the microscopic cause of resistance and the resulting generation of heat.}
	\label{fig:figure2}
	\end{figure}

But if there are no interactions in the channel, then how can heat be generated? The answer is that the heat is generated within the contacts (Fig.~\ref{fig:figure2}a)  which is not surprising since a bullet heats up the material it hits, and not the medium it flies through.  A corollary of this observation is that although any resistance $R$ is accompanied by the generation of $I^2R$ heat, the heating need not occur exactly where the resistance is caused.

For example in the ballistic conductor\index{ballistic conductor} the resistance is caused by the  channel-contact interfaces, while the heating could occur away from the interface. This separation of the resistance $R$ and its heat $I^2R$ is even more stark if we consider a ballistic conductor with a hole in the middle (Fig.~\ref{fig:figure2}b). We would expect the hole to cause a large increase in resistance by obstructing the flow of electrons. But it does not have the \emph{internal degrees of freedom}\index{internal degrees of freedom} needed to get heated which must still occur far away somewhere in the contacts\begin{footnote}{An experimental observation that supports this viewpoint is current transport through carbon nanotubes. Despite being extremely small in diameter and volume, carbon nanotubes can carry enormous amounts of current and if there were any significant energy dissipation within the channel, they could not get rid of such excessive heat and would burn up.}\end{footnote}.

However, pre-1990 it was not common to talk of resistance in a model without explicitly including interactions\index{interactions} and dissipation in the channel\begin{footnote}{Though device engineers had started to recognize that momentum and energy relaxation lengths could be very different in late 80's. See for example, Ref.~\cite{fischetti1988monte}.}\end{footnote}. Indeed much of the work on quantum transport was based on the Kubo formula which equates resistance to dissipation and then relates it to noise through the fluctuation-dissipation theorem. This is a powerful approach to linear response problems but it tends to create the impression that the origin of resistance is energy  relaxation rather than momentum relaxation. But it now seems clear that resistance\index{resistance} is caused by momentum exchange\index{momentum exchange}, and the resulting energy exchange need not occur at the same location. It is possible to create a zero-order model for resistance without including any $\pmb{\mathbf{\Sigma}}_{0}$ or $\pmb{\mathbf{\Sigma}}^{\text{in}}_{0}$ and this approach at least has enormous pedagogical value.

The NEGF Eqs.~(\ref{eq:eq1}-\ref{eq:eq3}) obtained in the next section from elementary quantum mechanics are the \textit{same} as those obtained in the classic paper by Keldysh using MBPT \cite{keldysh1965diagram}. The difference is that we obtain them considering only the physical contacts\index{contact} described by $\pmb{\mathbf{\Sigma}}_{1,2}$ and $\pmb{\mathbf{\Sigma}}^{\text{in}}_{1,2}$, while the classic treatment considers only the many-body interactions\index{many-body interactions} described by $\pmb{\mathbf{\Sigma}}_{0}$ and $\pmb{\mathbf{\Sigma}}^{\text{in}}_{0}$. Our approach makes it accessible to a much wider audience and for over a decade we have been teaching the NEGF equations to advanced undergraduate and beginning graduate students, both on-campus and online.

We believe the utility of this approach goes beyond the purely pedagogical. It decouples the NEGF\index{NEGF} Eqs.~(\ref{eq:eq1}-\ref{eq:eq3}) from MBPT\index{MBPT} thus clearing the path for alternative approaches to $\pmb{\mathbf{\Sigma}}_{0}$ and $\pmb{\mathbf{\Sigma}}^{\text{in}}_{0}$ including non-perturbative methods\index{non-perturbative methods} as well as phenomenological approaches. In this chapter we will present a simple example of the latter. 

\subsection{Outline}
In Sect.~\ref{sec:negf_summary} we will briefly summarize the NEGF equations followed by a brief outline of their derivation from the one-electron Schr\"{o}dinger equation\index{one-electron Schr\"{o}dinger equation} in Sect.~\ref{sec:negf_derivation}. In Sect.~\ref{sec:example} we present toy examples to give the reader a feeling for how the method is applied. Corresponding Python codes are also included for interested readers. We end in Sect.~\ref{sec:FAQ} with answers to a few commonly asked questions. For more detailed examples and discussions we refer the reader to Chapter 8 of Ref.~\cite{datta1997electronic}, Ref.~\cite{datta2005quantum} and Part B of Ref.~\cite{datta2012lessons}.

\section{NEGF Equations}
\label{sec:negf_summary}
There are three NEGF equations\index{NEGF equations} as described below. \\

\begin{enumerate}
\item First is the equation for the quantum \textit{density of states}\index{quantum density of states} (times 2$\pi$)\begin{footnote}{Note that the quantum density of states is generally a dense matrix of size $ N \times N$ for a lattice size of $N$ ($2N \times 2N$ if electron spin is included). Its diagonal entries correspond to the classical local density of states but off-diagonal elements also contain useful information. For example, when the quantum density of states is multiplied by \index{Pauli spin matrices}Pauli spin matrices, the diagonal elements provide local density of states \emph{for a particular spin} as we show later by an example in Section~\ref{sec:diffspin}. The matrix nature of $\mathbf{A}$ and $\mathbf{G}^n$ compared to their classical counterparts that are of size $N \times 1$ is due to the quantum generalization of these quantities.}\end{footnote}
\begin{equation*}
\mathbf{A}= i\left[\mathbf{G}^{R} - \mathbf{G}^{A}\right]
\end{equation*}
where the advanced Green function $\mathbf{G}^{A}$ is the Hermitian conjugate of the retarded Green function $\mathbf{G}^{R}$ given by $(\pmb{\mathbf{\Sigma}}=\pmb{\mathbf{\Sigma}}_0+\pmb{\mathbf{\Sigma}}_1+\pmb{\mathbf{\Sigma}}_2)$
\begin{equation} 
\mathbf{G}^{R} \; =\; \left[E\mathbf{I}-\mathbf{H}-\pmb{\mathbf{\Sigma}}\right]^{-1}
\label{eq:eq1}  
\end{equation}

\item Next is the equation for the quantum \textit{electron density}\index{electron density} (times 2$\pi$) per unit energy
\begin{equation} 
\mathbf{G}^{n} \; =\; \mathbf{G}^{R} \,  \pmb{\mathbf{\Sigma}}^{\text{in}}\, \mathbf{G}^{A}
\label{eq:eq2}  
\end{equation} 

\item And finally the current\index{current} per unit energy $\tilde{I}_m$ at contact $m$ is given by
\begin{equation} 
\tilde{I}_{m} \;=\; \frac{q}{h} \, \mbox{Trace}\,\left[\pmb{\mathbf{\Sigma}}_{m}^{\text{in}} \mathbf{A}-\mathbf{\Gamma}_{m} \, \mathbf{G}^{n}\right]
\label{eq:eq3} 
\end{equation} 
\noindent where \  $ \mathbf{\Gamma_m} =i\left[\pmb{\mathbf{\Sigma}}_m -\pmb{\mathbf{\Sigma}}_m^{\dagger} \right]$.
\end{enumerate}

We use a notation that is slightly different from that used in the conventional literature because it helps provide a physical picture for different quantities. For example we use $\mathbf{G}^{n}$ instead of the traditional $-i\mathbf{G}^{<}$ because $\mathbf{G}^{n}$ represents a matrix version of the \emph{electron density}\index{electron density}. The correspondences are summarized in a table to help the reader translate our equations as needed.
\begin{table}[!ht]
\centering
\caption{NEGF symbols used in this chapter and their counterpart in the literature.}
\begin{tabular}{c@{\hspace{1cm}}c@{\hspace{1cm}}c}
\hline\noalign{\smallskip}
		\textbf{Conventional} & \textbf{Physical} & \textbf{Symbol used}\\
		\textbf{symbol}  & \textbf{Interpretation} & \textbf{in this book}\\
		\noalign{\smallskip}\svhline\noalign{\smallskip}
		$-i\mathbf{G}^{<}$     & Matrix electron density & $\mathbf{G}^{n}$\\ 
		$+i\mathbf{G}^{>}$     & Matrix hole density & $\mathbf{G}^{p}$\\ 
		$-i\pmb{\mathbf{\Sigma}}^{<}$ & In-scattering function & $\pmb{\mathbf{\Sigma}}^{\text{in}}$\\
		$+i\pmb{\mathbf{\Sigma}}^{>}$ & Out-scattering function & $\pmb{\mathbf{\Sigma}}^{\text{out}}$\\
		\noalign{\smallskip}\hline\noalign{\smallskip}
\end{tabular}
\label{tab:tab1}
\end{table}

\section{NEGF equations from one-electron Schr\"{o}dinger equation} \index{one-electron Schr\"{o}dinger equation}
\label{sec:negf_derivation}

Let us briefly indicate how the NEGF Eqs.~(\ref{eq:eq1}-\ref{eq:eq3}) are obtained  directly from the one-electron Schr\"{o}dinger equation  $E\, \pmb{\mathbf{\psi}} \; =\; \mathbf{H}\, \pmb{\mathbf{\psi}} $. The first step is to incorporate the effect of the \emph{open boundary conditions}\index{open boundary conditions} imposed by the contacts through two additional terms 

\begin{equation}
E\, \pmb{\mathbf{\psi}} \; =\; \mathbf{H}\, \pmb{\mathbf{\psi}}  \; + \; \underbrace{\pmb{\mathbf{\Sigma}}\, \pmb{\mathbf{\psi}}}_{\text{OUTFLOW}}\;  + \; \underbrace{\mathbf{s} }_{\text{INFLOW}}
\label{eq:eq4}
\end{equation}

Note that the modification of the Schr\"{o}dinger equation by the $\pmb{\mathbf{\Sigma}}$ and the $\mathbf{s}$ terms lead to a conceptually different viewpoint of energy. Instead of being the resonant energies of a given system described by $\mathbf{H}$, energy now becomes an independent variable related to the incoming electrons from the contacts that excite the channel. (see Chapter 8 of \cite{datta2005quantum} for a detailed discussion).  

\subsection{NEGF Eqs.~(\ref{eq:eq1}-\ref{eq:eq2})}

From Eq.~(\ref{eq:eq4}) we write the wavefunction\index{wavefunction} as
\[\pmb{\mathbf{\psi}}\; = \; \left[E\mathbf{I}-\mathbf{H}-\pmb{\mathbf{\Sigma}}\right]^{-1}\mathbf{s} \;=\; \mathbf{G}^{R} \, \mathbf{s} \]
\noindent making use of the definition of $\mathbf{G}^{R}$ from NEGF Eq.~(\ref{eq:eq1}). Note that $\pmb{\mathbf{\Sigma}}$ is non-Hermitian so that $E\mathbf{I}-\mathbf{H}-\pmb{\mathbf{\Sigma}}$ is not singular for any real value of $E$ and has a well-defined inverse $\mathbf{G}^R$.

Since different sources `$\mathbf{s}$' are incoherent, we cannot superpose the resulting $\pmb{\mathbf{\psi}}$'s from multiple sources. So we define bilinear quantities that \emph{can} be superposed:
\[\underbrace{\pmb{\mathbf{\psi}} \pmb{\mathbf{\psi}}^{\dagger} }_{\mathbf{G}^{n}/(2\pi) } \; = \; \mathbf{G}^{R}\, \underbrace{\mathbf{s} \, \mathbf{s}^{\dagger} }_{\pmb{\mathbf{\Sigma}} ^{\text{in}}/(2\pi) }\, \mathbf{G}^{A}\]
giving us NEGF Eq.~(\ref{eq:eq2}).

In our description $\pmb{\mathbf{\psi}}$ is the one-electron wavefunction whose square gives the probability of finding an electron. When averaged over all electrons in an ensemble it gives the electron density. More generally $\pmb{\mathbf{\psi}}\pmb{\mathbf{\psi}}^{\dagger}$ is a matrix whose diagonal elements give the probabilities, and the off-diagonal elements give the correlations. For a more formal justification using field operators the reader can check the Appendix in Ref.~\cite{datta2005quantum}.

Note that $\mathbf{s} \mathbf{s}^{\dagger}$ reflects the availability of electrons at a given energy in the contacts which fill the available states in the channel giving rise to $\pmb{\mathbf{\psi}}\pmb{\mathbf{\psi}}^{\dagger}$. Fermi functions are invoked only in the contacts and not in the channel.

\subsection{NEGF Eq.~(\ref{eq:eq3})}
Starting from the time-dependent version of the modified Schr\"{o}dinger equation, Eq.~(\ref{eq:eq4}), 
\[i\hbar \, \frac{d}{dt} \,\pmb{\mathbf{\psi}} \; =\; [\mathbf{H+\Sigma}]\, \pmb{\mathbf{\psi}}  \; +\;  \mathbf{s} \] 
and its conjugate transpose (noting that $\mathbf{H}$ is a Hermitian matrix\index{Hermitian matrix})
\[-\, i\hbar \, \frac{d}{dt} \, \pmb{\mathbf{\psi}}^{\dagger} \; =\; \pmb{\mathbf{\psi}}^{\dagger} \, [\mathbf{H+\Sigma}^{\dagger} ]\;  +\;  \mathbf{s}^{\dagger} \] 
and making use of the relations
\[\pmb{\mathbf{\psi}}\;=\;\mathbf{G}^{R}\,\mathbf{s} \quad \mbox{and} \quad \pmb{\mathbf{\psi}}^{\dagger} \; =\; \mathbf{s}^{\dagger}\,\mathbf{G}^{A},\] 
we can write
\[\begin{split}
i\hbar \, \dfrac{d}{dt} \pmb{\mathbf{\psi}} \pmb{\mathbf{\psi}} ^{\dagger}  &= \left(i\hbar \, \frac{d}{dt} \, \pmb{\mathbf{\psi}} \right)\, \pmb{\mathbf{\psi}} ^{\dagger} \; +\; \pmb{\mathbf{\psi}} \, \left(i\hbar \, \dfrac{d}{dt} \, \pmb{\mathbf{\psi}} ^{\dagger} \right)\\
&= \left([\mathbf{H+\Sigma}]\, \pmb{\mathbf{\psi}}  + \mathbf{s} \right) \pmb{\mathbf{\psi}}^{\dagger} - \pmb{\mathbf{\psi}} \left(\pmb{\mathbf{\psi}}^{\dagger}\,[\mathbf{H+\Sigma^{\dagger}} ] + \mathbf{ s}^{\dagger} \right)\\
&= \left[(\mathbf{H+\Sigma})\, \pmb{\mathbf{\psi}} \pmb{\mathbf{\psi}} ^{\dagger} -\pmb{\mathbf{\psi}} \pmb{\mathbf{\psi}} ^{\dagger} (\mathbf{H+\Sigma} ^{\dagger} )\right]\;+\; \left[\mathbf{s}\mathbf{s}^{\dagger} \mathbf{G}^{A} -\mathbf{G}^{R} \mathbf{s}\mathbf{s}^{\dagger}\right]
\end{split}\] 
Making use of
\[2\pi \, \pmb{\mathbf{\psi}} \pmb{\mathbf{\psi}} ^{\dagger} \;=\; \mathbf{G}^{n} \quad    \mbox{and} \quad 2\pi \, \mathbf{s} \mathbf{s} ^{\dagger}\;=\;\mathbf{\Sigma_{\text{in}}}\]
we can write
\begin{equation}
\dfrac{d}{dt} \pmb{\mathbf{\psi}} \pmb{\mathbf{\psi}} ^{\dagger}  \;=\; \frac{[\mathbf{HG}^{n} -\mathbf{G}^{n}\mathbf{H}]\; +\; [\mathbf{\Sigma G}^{n} -\mathbf{G}^{n}\pmb{\mathbf{\Sigma}}^{\dagger}]\;+\; [\pmb{\mathbf{\Sigma}}^{\text{in}}\mathbf{G}^{A}-\mathbf{G}^{R}\pmb{\mathbf{\Sigma}}^{\text{in}}]}{i\, 2\pi\hbar}. 
\label{eq:eq5}
\end{equation} 

\noindent We can interpret  $\pmb{\mathbf{\psi}}\pmb{\mathbf{\psi}}^{\dagger}$  as the number operator\index{number operator} so that its derivative represents the current operator\index{current operator}. To find the change of any quantity $A$, we can multiply by the corresponding operator $\mathbf{A}_{\text{op}}$ and take the trace to obtain 
\[\frac{dA}{dt} =-\frac{i}{h} \mbox{Trace} \left( [\mathbf{H G}^{n} -\mathbf{G}^{n} \pmb{\mathbf{H}} ] \mathbf{A}_{\text{op}} +  [\mathbf{\Sigma G}^{n} -\mathbf{G}^{n} \pmb{\mathbf{\Sigma}}^{\dagger} ] \mathbf{A}_{\text{op}} + [\pmb{\mathbf{\Sigma}}^{\text{in}} \mathbf{G}^{A} -\mathbf{G}^{R} \pmb{\mathbf{\Sigma}}^{\text{in}} ] \mathbf{A}_{\text{op}} \right)\] 

\noindent The first term represents the change in the quantity $A$ inside the channel due to the action of the Hamiltonian $\bf{H}$ while the last two terms represent the ``$A$ \emph{current}''  injected from the terminals. We can define the terminal current operator $\mathbf{I}_{\text{op}}$ as

\begin{equation}
\mathbf{I}_{\text{op}} =-\frac{i}{h} \; \, \left(\,  [\mathbf{\Sigma G}^{n} -\mathbf{G}^{n} \pmb{\mathbf{\Sigma}}^{\dagger} ]\; +\; \, [\pmb{\mathbf{\Sigma}}^{\text{in}} \mathbf{G}^{A} -\mathbf{G}^{R} \pmb{\mathbf{\Sigma}}^{\text{in}} ]\; \right)
\label{eq:currentOp}
\end{equation}
such that the ``$A$ \emph{current}'' can be obtained from the trace of $\mathbf{A}_{\text{op}}\mathbf{I}_{\text{op}}$.

\noindent If we are interested only in the charge current, then we need the trace of $\mathbf{I}_{\text{op}}$ since the charge operator is an identity matrix (times the electronic charge). Noting that $\mbox{Trace}\,[\mathbf{A}\mathbf{B}] = \mbox{Trace}\, [\mathbf{B}\mathbf{A}]$ we can write
\[ \mbox{Trace} \left(\mathbf{I}_{\text{op}}\right) \;=\; \dfrac{1}{h}\, \mbox{Trace} \left(\pmb{\mathbf{\Sigma}}^{\text{in}}\mathbf{A} -\mathbf{\Gamma}\mathbf{G}^{n} \right).\] 

\noindent Both the left and the right hand sides of this equation are zero, since we are discussing steady state charge transport with no time variation. But the terms on the left can be separated into two parts, one associated with contact 1 and one with contact 2. They tell us the currents at contacts 1 and 2 respectively and the fact that they add up to zero is simply a statement of \emph{Kirchhoff's law}\index{Kirchhoff's law} for steady-state currents\index{steady-state currents} in circuits\begin{footnote}{Note that Eq.~(\ref{eq:currentOp}) can also be used to obtain terminal \emph{spin} currents if both sides of the equation is multiplied by Pauli spin matrices\index{Pauli spin matrices} corresponding to a desired spin direction. In this case however the standard Kirchhoff's Laws for charge transport may not hold but it is still possible to design ``spin-circuits''\index{spin-circuits} that obey Kirchhoff's laws \cite{camsari2019non}.}\end{footnote}.

\noindent With this in mind we can write for the current at contact $m$ ($m=1,2$)
\[\tilde{I}_{m} \;=\; \frac{q}{h}\, \mbox{Trace} \left(\pmb{\mathbf{\Sigma}}_{m}^{\text{in}} \mathbf{A}\;-\; \mathbf{\Gamma}_{m} \, \mathbf{G}^{n} \right)\] 
\noindent as stated in NEGF Eq.~(\ref{eq:eq3}).
\section{A simple example}
\label{sec:example}
In this Section we will outline the application of the NEGF\index{NEGF} Eqs.~(\ref{eq:eq1}-\ref{eq:eq2}) to a simple  example centered around a one-dimensional conductor with a barrier $U$ and connected to two contacts with Fermi functions\index{Fermi fucntions} $f_1(E) = 1$ and $f_2(E)=0$.  The reader can consult Part B of Ref.~\cite{datta2012lessons} for more details. 
\begin{figure}[!ht]
\centering
	\includegraphics[width=0.99\linewidth,keepaspectratio]{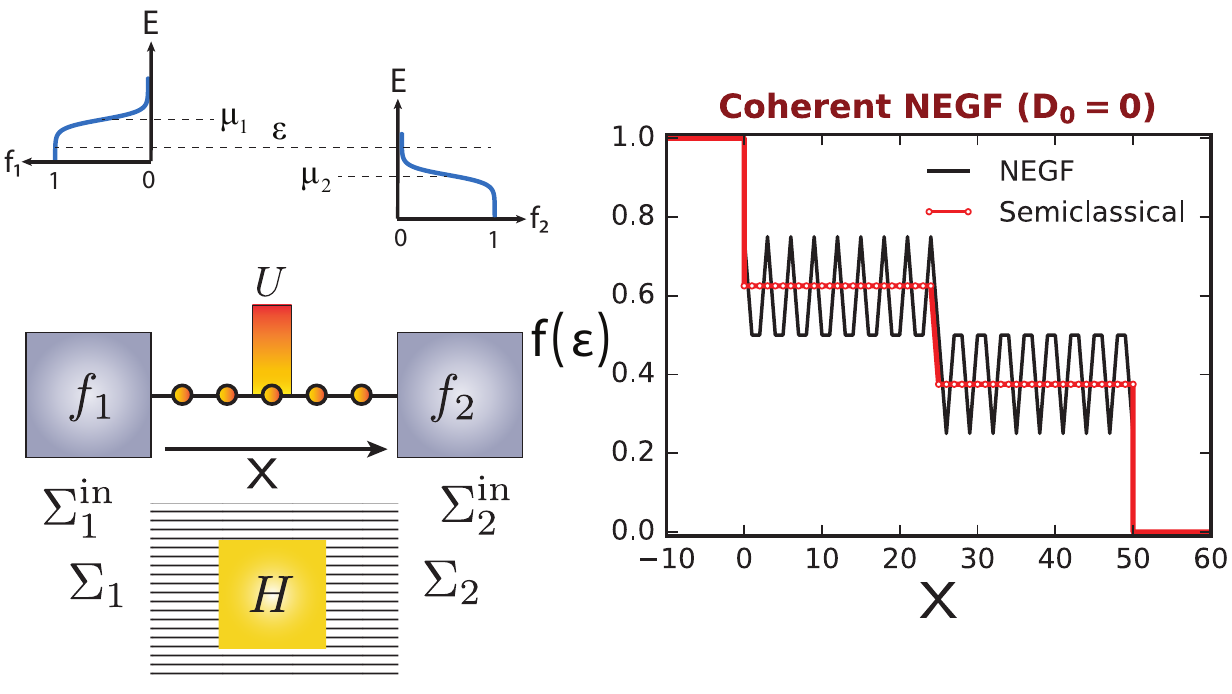}
	\caption{\textbf{Left-Top}: A simple one-dimensional conductor with a barrier $U$ and connected to two contacts with Fermi functions $f_1(E) = 1$ and $f_2(E)=0$. \textbf{Left-Bottom}: To analyze it  using the NEGF Eqs.~(\ref{eq:eq1}-\ref{eq:eq2}) we need the $\mathbf{H}$ to describe the channel and the $\pmb{\mathbf{\Sigma}}$'s to describe the physical contacts. Interactions in the channel are ignored in this example ($D_0$ = 0). \textbf{Right}: Plot of the occupation factor across the channel. }
	\label{fig:figure3}
	\end{figure}
\subsection{Hamiltonian  $\mathbf{H}$}
\begin{figure}[!ht]
\sidecaption
\includegraphics[width=0.55\linewidth,keepaspectratio]{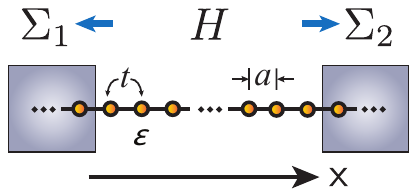}
\caption{For the one-dimensional examples discussed in this chapter, we use the 1D Hamiltonian  shown with on-site elements $\varepsilon$ and nearest neighbor coupling $t$.}
\label{fig:figure3a}
\end{figure}

We describe the 1D channel using a simple tight-binding Hamiltonian\index{tight-binding Hamiltonian} $\mathbf{H}$ with $\varepsilon$ on the diagonal and $t$ on the upper and lower diagonals, as shown pictorially in Fig.~\ref{fig:figure3a}. This leads to a dispersion relation of the form
\begin{subequations}
\begin{equation}\label{eq:eq10a}
E(k)\;=\;\varepsilon+2t\cos{(ka)}
\end{equation}
\noindent This cosine dispersion\index{cosine dispersion} can approximate a parabolic one
\begin{equation}\label{eq:eq10b}
E\;=\; E_{c} +\frac{\hbar ^{2} k^{2} }{2m}
\end{equation}
\label{eq:eq10}
\end{subequations}
if we choose
\begin{subequations}
\begin{equation}\label{eq:1eq1a}
E_{c} \;=\; \varepsilon +2t
\end{equation}
\begin{equation}\label{eq:eq11b}
\mbox{and } \quad -\, t\equiv t_{0} \equiv \dfrac{\hbar ^{2} }{2ma^{2} }.
\end{equation}
\label{eq:eq11}
\end{subequations}

\subsection{Self-energy due to contacts}
The self-energy function for each contact has only one non-zero element corresponding to the point that is connected to that contact:
\begin{equation*}
\pmb{\mathbf{\Sigma}} _{1} \;=\; \left[\begin{array}{C{0.6cm}C{0.6cm}C{0.6cm}C{0.6cm}C{0.6cm}} te^{ika} & 0 & 0 & \cdots & 0 \\ 0 & 0 & 0 & \cdots & 0 \\ 0 & 0 & 0 & \cdots & 0 \\ \vdots & \vdots & \vdots & \ddots & \vdots \\ 0 & 0 & 0 & \cdots & 0 \end{array}\right],
\quad 
\pmb{\mathbf{\Sigma}}_{2} \;=\; \left[\begin{array}{C{0.6cm}C{0.6cm}C{0.6cm}C{0.6cm}C{0.6cm}} 0 & \cdots & 0 & 0 & 0 \\ \vdots & \ddots & \vdots & \vdots & \vdots \\ 0 & \cdots & 0 & 0 & 0 \\ 0 & \cdots & 0 & 0 & 0 \\ 0 & \cdots & 0 & 0 & te^{ika} \end{array}\right]
\end{equation*} 

\subsection{Inscattering from contacts}
The inscattering functions\index{inscattering functions} are given simply by the broadening functions\index{broadening functions} times the corresponding occupation factor\index{occupation factor}:
\begin{equation}
\pmb{\mathbf{\Sigma}}_1^{\text{in}} = \mathbf{\Gamma}_1 \ f_1 \quad \mathrm{and} \quad \pmb{\mathbf{\Sigma}}_2^{\text{in}} = \mathbf{\Gamma}_2 \ f_2
\end{equation}

\noindent where the broadening functions\index{broadening functions} are obtained from $\mathbf{\Gamma}_{1,2}=i\,[\pmb{\mathbf{\Sigma}}_{1,2} - \pmb{\mathbf{\Sigma}}_{1,2}^{\dagger}]$:
\begin{equation*}
\mathbf{\Gamma _{1}} \;=\; \frac{\hbar \nu}{a}\left[\begin{array}{C{0.6cm}C{0.6cm}C{0.6cm}C{0.6cm}C{0.6cm}} 1 & 0 & 0 & \cdots & 0 \\ 0 & 0 & 0 & \cdots & 0 \\ 0 & 0 & 0 & \cdots & 0 \\ \vdots & \vdots & \vdots & \ddots & \vdots \\ 0 & 0 & 0 & \cdots & 0 \end{array}\right],
\quad 
\mathbf{\Gamma_{2}} \;=\; \frac{\hbar \nu}{a} \left[\begin{array}{C{0.6cm}C{0.6cm}C{0.6cm}C{0.6cm}C{0.6cm}} 0 & \cdots & 0 & 0 & 0 \\ \vdots & \ddots & \vdots & \vdots & \vdots \\ 0 & \cdots & 0 & 0 & 0 \\ 0 & \cdots & 0 & 0 & 0 \\ 0 & \cdots & 0 & 0 & 1 \end{array}\right]
\end{equation*}

\noindent noting that the velocity\index{velocity} $\nu = dE/(\hbar dk)$ which  from the dispersion relation\index{dispersion relation} Eq.~(\ref{eq:eq10a}) equals $ -2a t / \hbar \sin{(ka)} $.\\

Given these matrices it is straightforward to use the NEGF\index{NEGF} Eqs.~(\ref{eq:eq1}-\ref{eq:eq2}) to evaluate the ``electron density''\index{electron density} $\mathbf{G}^{n}/2 \pi$ and the ``density of states''\index{density of states} $\mathbf{A}/2 \pi$ from which we can obtain the $\ell$-th element of the occupation factor
\begin{equation*}
f_{\ell} = \frac{G^n_{\ell,\ell}}{A_{\ell,\ell}}
\end{equation*}
which is plotted in Fig.~\ref{fig:figure3}.  Note that the NEGF occupation factor shows oscillations due to quantum interference\index{quantum interference} around the average values expected from a semiclassical picture. The semiclassical curve is based on three localized resistors\index{localized resistors} in series, one each at the two interfaces\index{interface}, and one at the barrier: 
\begin{equation} 
\frac{R}{h/q^2} = \underbrace{\frac{1}{2}}_{\text{channel-source interface}}+  \underbrace{\frac{1-T}{T}}_{\text{barrier}} +  \underbrace{\frac{1}{2}}_{\text{channel-drain interface}}
\label{eq:semiclassical_R}
\end{equation}

\noindent where $T$ is the transmission probability\index{transmission probability} through the barrier given by
\begin{equation}
T(E)  \;=\; \dfrac{(\hbar \nu /a)^{2} }{U^{2} +(\hbar \nu /a)^{2} }  
\label{eq:tr}
\end{equation}
This is a standard result that can be obtained from scattering theory on a discrete lattice \cite{datta2005quantum}, next we will show how the NEGF equations lead to the same result. 

\subsection{Current}
If we are not interested in the spatial variation of quantities like the occupation factor, but are only interested in the current that flows for a given voltage, we can simply let the channel represent one point, and treat the rest as contacts. This makes all matrices collapse to (1 $\times$ 1) reducing them to just numbers:

\begin{equation*} 
H = \varepsilon + U
\end{equation*}
\begin{equation*} 
{\Sigma}_1 (E) = t e^{+ika}  = {\Sigma}_2 (E)  \quad \text{where } ka = \cos^{-1}{ \bigg( \frac{E - \varepsilon } {2t} \bigg)}
\end{equation*}
\begin{equation*} 
{\Sigma}(E)   = {\Sigma}_1 (E)  + {\Sigma}_2 (E) = 2t e^{+ika}
\end{equation*}
\begin{equation*} 
\Gamma_1 = \frac{\hbar \nu}{a}  = \Gamma_2
\end{equation*}
\begin{equation*} 
{\Sigma}_1^{\text{in}} = \Gamma_1 f_1, \quad {\Sigma}_2^{\text{in}}   = \Gamma_2 f_2 \quad \text{and} \quad {\Sigma}^{\text{in}} = \Gamma_1 f_1 + \Gamma_2 f_2
\end{equation*}
\noindent Now we do not need a computer to evaluate the NEGF Eqs.~(\ref{eq:eq1}-\ref{eq:eq3}). We can just do it by hand. First we calculate $G^{R}$ from Eq.~(\ref{eq:eq1}):
\[{G}^{R} (E)\; =\; \dfrac{1}{E-(\varepsilon +U)-2t\, e^{ika} } \]
\[ \quad \quad \quad \quad \quad \quad \quad \quad \quad \; =\; \dfrac{1}{-\,U-i2t\, \sin{(ka)}}       \; =\; \dfrac{1}{-\,U + i \hbar \nu/a\,  { }}                         \] 
\noindent so that
\begin{equation*}
{G}^{A} (E) \; =\; \dfrac{1}{-\,U - i \hbar \nu/a\, { }}
\end{equation*}
\begin{equation*}
{A}(E) \; =\; \dfrac{2 \hbar \nu / a}{U^2+ (\hbar \nu/a)^2 \, { }}
\end{equation*}
\noindent Next from Eq.~(\ref{eq:eq2}):
\begin{equation*}
G^{n} =  \frac{\hbar \nu/a}{U^2 + (\hbar \nu/a)^2} \big( f_1 + f_2 \big)
\end{equation*}
\noindent so that from Eq.~(\ref{eq:eq3}):
\begin{equation*}
\tilde{I}(E) = \frac{q}{h} \frac{ (\hbar \nu/a)^2 }{U^2 + (\hbar \nu/a)^2} \big( 2f_1 - \overline{f_1+ f_2}  \big)
\end{equation*}
\begin{equation*}
 = \frac{q}{h}  \ \   \underbrace {\frac{ (\hbar \nu/a)^2 }{U^2 + (\hbar \nu/a)^2} }_{T(E)} \big( f_1 - f_2  \big)
\end{equation*}
\noindent The quantity $T(E)$ represents the transmission probability\index{transmission probability} through the barrier $U$. Integrating $\tilde{I}$(E) we get the total current\index{current}. The quantity $T(E)$ is the same as the semiclassical transmission that was shown in Eq.~(\ref{eq:tr}). 

\subsection{Dephasing interactions}

\begin{figure}[!ht]
\centering
	\includegraphics[width=\linewidth,keepaspectratio]{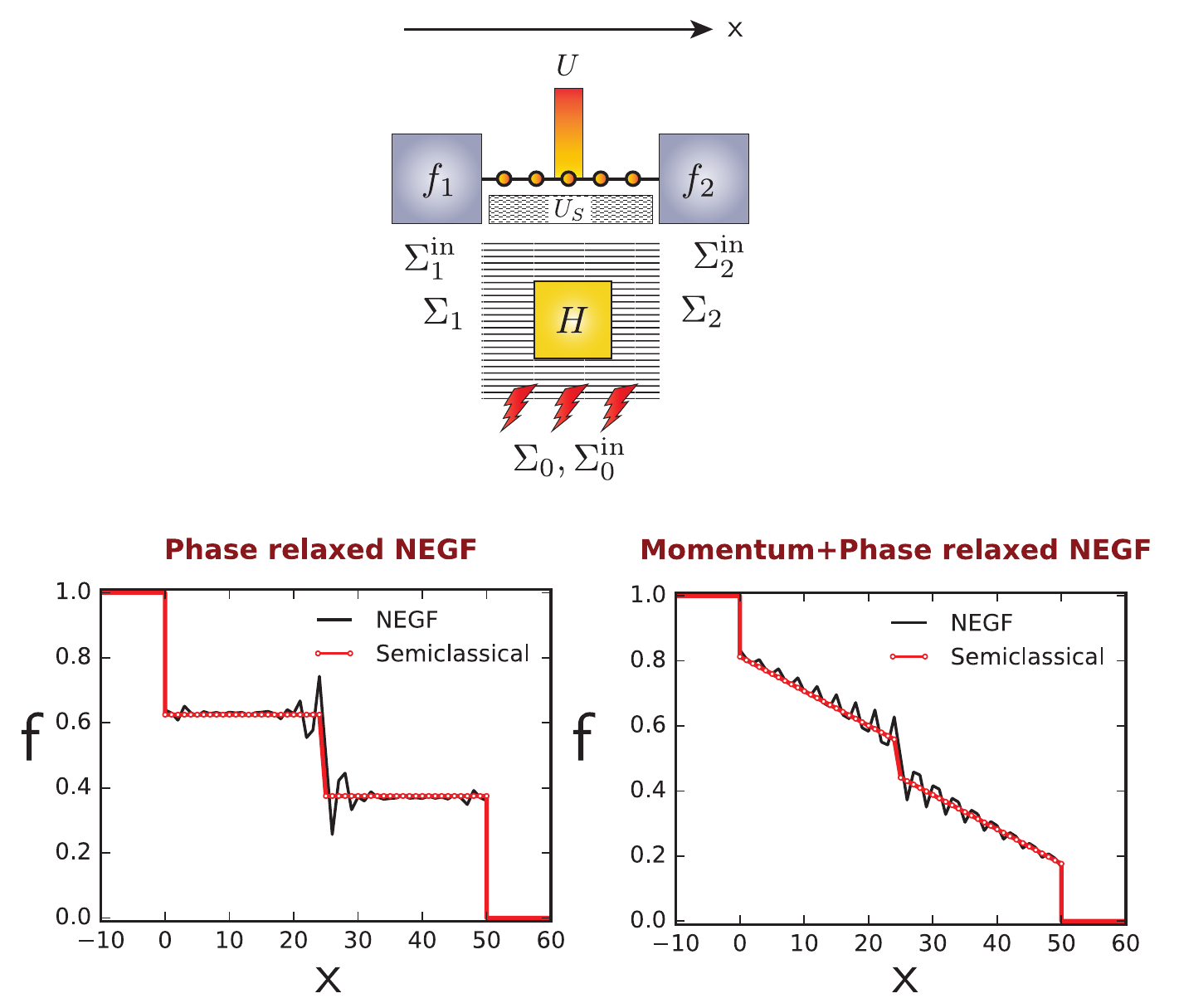} 
	\caption{\textbf{Top:} Same problem as in Fig.~\ref{fig:figure3} but we now include the effect of a random potential\index{random potential} $U_{S}$. \textbf{Middle:} The effect of the random potential is incorporated through  $\pmb{\mathbf{\Sigma}}_{0}$ and $\pmb{\mathbf{\Sigma}}^{\text{in}}_{0}$ where the magnitude of scattering, $D_0$, is 0.09 $t_0^2$. \textbf {Bottom Left:} Plot of occupation factor\index{occupation factor} across channel for phase-relaxing scattering. \textbf{Bottom right:} Plot of occupation factor across channel for phase and momentum-relaxing scattering\index{phase-relaxing scattering}\index{momentum-relaxing scattering}.}
	\label{fig:figure4}
	\end{figure}

What we have done so far represents coherent transport\index{coherent transport} with no interactions in the channel. In this limit, the results could also have been obtained from the scattering theory of transport, sometimes referred to as the Landauer-B\"{u}ttiker formalism\index{Landauer-B\"{u}ttiker formalism}. The NEGF method provides a convenient way to get the same results. However, the NEGF\index{NEGF} method also allows us to include channel interactions through the self-energy functions\index{self-energy function} $\pmb{\mathbf{\Sigma}}_{0}$, $\pmb{\mathbf{\Sigma}}^{\text{in}}_{0}$.

It is important to include these interactions because the purely coherent model predicts quantum interference\index{quantum interference} effects like the oscillations in Fig.~\ref{fig:figure3} which are usually not seen experimentally except at extremely low temperatures. The reason is that the phase of individual electrons is destroyed by the random scattering potential\index{scattering potential} it sees due to the other electrons. Such electron-electron interactions\index{electron-electron interactions} normally do not lead to any overall loss of momentum from the system of electrons and so have little effect on the mobility or the current. But they do lead to dephasing\index{dephasing} which suppresses the coherent oscillations. Dephasing may also arise from other sources such as electron-phonon interactions \cite{ALTSHULLER1981619}. However, they usually cause momentum relaxation as well, thereby affecting the resistance. We specifically mention electron-electron interaction because it can cause dephasing without having any first-order effect on the resistance. Resistance typically becomes temperature independent around 10 K, once phonons have frozen out. But electron-electron interactions continue to cause dephasing and it usually takes much lower temperatures to see interference effects.

The classic NEGF provides definite prescriptions for including all such effects from first principles starting from a microscopic Hamiltonian using MBPT\index{MBPT} as described in many available references. Here we will use a phenomenological approach that incorporates dephasing in terms of a single parameter $D_0$ which can be adjusted to reflect experimental dephasing times:

\begin{subequations}
\begin{equation}
\pmb{\mathbf{\Sigma}}_{0} \; =\; D_{0} \  \mathbf{G}^{R}
\end{equation}
\begin{equation}
\pmb{\mathbf{\Sigma}}_{0}^{\text{in}} \; =\; D_{0} \ \mathbf{G}^{n}
\end{equation}
\end{subequations}

\noindent Note that we now need to solve Eqs.~(\ref{eq:eq1}) and (\ref{eq:eq2}) self-consistently since the same quantity appears on both the left and right hand sides:
\begin{subequations}
\begin{equation} 
\mathbf{G}^{R} \; =\; \left[E\mathbf{I}-\mathbf{H}-\pmb{\mathbf{\Sigma_1}} -\pmb{\mathbf{\Sigma_2}}  - D_0 \mathbf{G^{R}}   \right]^{-1}
\end{equation}
\begin{equation}
\mathbf{G}^{n} \; =\; \mathbf{G}^{R} \, [  \pmb{\mathbf{\Sigma}}_1^{\text{in}}\, + \pmb{\mathbf{\Sigma}}_2^{\text{in}}\, + D_0 \mathbf{G}^{n}\,] \mathbf{G}^{A}
\label{eq:eq20}  
\end{equation}
\end{subequations}

The self-consistent calculation\index{self-consistent calculation} is straightforward to implement numerically and a Python code is provided for readers interested in reproducing the results shown in Figs.~\ref{fig:figure3} and \ref{fig:figure4}. Note that the results for pure dephasing agree well with the simple semiclassical model based on series resistors (Eq. \ref{eq:semiclassical_R}) as described earlier. Fig. \ref{fig:figure4} also shows results for a slightly different model of interactions which relaxes both phase and momentum leading to a linear drop in the occupation factor across the channel like an ohmic resistor. In this slightly different model, the diagonal elements of  $\pmb{\mathbf{\Sigma}}_{0}$ and $\pmb{\mathbf{\Sigma}}^{\text{in}}_{0}$ are the same as before, but all off-diagonal elements are set to zero \cite{golizadeh2007nonequilibrium,datta2008nanoelectronic}.

Intuitively, we can justify the pure phase relaxation model by thinking of it as distributed ``B\"{u}ttiker probes'' that take electrons out of the channel and reinject them back into it to break their phase coherence \cite{hershfield1991equivalence}. But because the electron density ($\mathbf{G}^n$)  is preserved \textit{exactly} by constant multiplication, no  property other than phase is relaxed. In contrast, just keeping the diagonal elements of the $\mathbf{G}^n$ in the real space representation breaks momentum in addition to phase since the reinjected electrons have no definite momentum and can flow in any direction. This intuition allows an immediate extension to spin preserving but momentum relaxing $\pmb{\mathbf{\Sigma}}_0$ as follows: Supposing we express $\mathbf{H}$, $\mathbf{G}^n$ and lead $\pmb{\mathbf{\Sigma}}$'s  in $2N \times 2N$ matrices for a lattice of $N$ points, following the same principle we could write down a spin-conserving but momentum and phase relaxing $\pmb{\mathbf{\Sigma}}_0$ as: 

\begin{equation*}
\pmb{\mathbf{\Sigma}}_{0} \;=\; \left[\begin{array}{C{0.6cm}C{0.6cm}C{0.6cm}C{0.6cm}C{0.6cm}C{0.6cm}} \begin{bmatrix}1 & 1 \\ 1 & 1 \end{bmatrix} & 0 & \cdots & \cdots  & 0  \\ 0 &  \begin{bmatrix}1 & 1 \\ 1 & 1 \end{bmatrix} & 0 & \cdots & 0   \\ \vdots & \vdots & \vdots & \ddots & \vdots  \\   0 & 0 & \cdots & \cdots & \begin{bmatrix}1 & 1 \\ 1 & 1 \end{bmatrix}  \end{array} \ \right],
\end{equation*}
where the $2\times 2$ blocks in each lattice point preserve spin information but phase and momentum are relaxed. Indeed, using this principle we can write down $\pmb{\mathbf{\Sigma}}_0$'s that can preserve information within \textit{any} unit cell with $B$ basis functions by having $B\times B$ blocks of all 1's that preserve local correlations in $\mathbf{\mathbf{G}}^n$. In the next subsection, we show an example of spin conserving dephasing in the context of spin transport. 
\begin{figure}[!t]
\centering
	\includegraphics[width=0.95\linewidth,keepaspectratio]{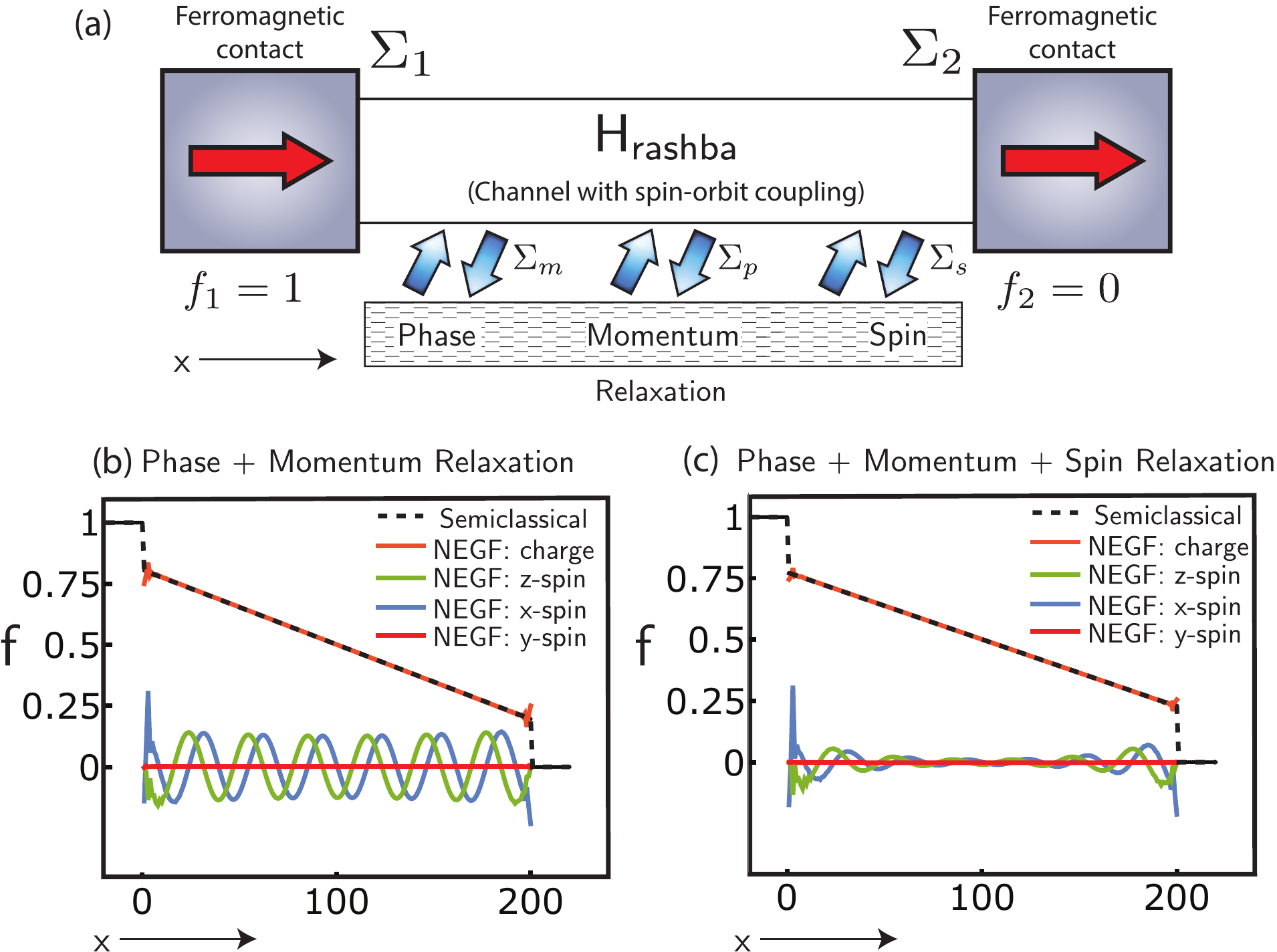}
	\caption{(a) Rashba channel with ferromagnetic leads. Phase, momentum and spin relaxation can be introduced to the system independently. (b) Spin precession is possible when $+x$ spins are injected to the channel even in the presence of phase and momentum relaxation. (c) Adding spin relaxation attenuates spin precession. (Python code available at the end of the article).}
	\label{fig:rashba}
	\end{figure}

\subsection{Diffusive spin transport using NEGF} 
\label{sec:diffspin}
Spin electronics or spintronics that combine spin transport with magnetics has attracted much attention in recent years \cite{vzutic2004spintronics}.  Of particular interest are the so-called high spin orbit materials such as Rashba channels or  interfaces \cite{sanchez2013spin,caviglia2010tunable,lesne2016highly,jungfleisch2018control}  as well as topological insulators \cite{tian2015electrical} or heavy metals with high spin orbit coupling \cite{miron2010current,liu2012spin}.  NEGF allows a natural extension to incorporate spin transport in the tight-binding framework which makes the Hamiltonian for a lattice of $N$ points a $2N \times 2N$ matrix. For example, consider a channel with a Rashba spin-orbit interaction that is described by the Hamiltonian, 
\begin{equation}
\mathbf{H}= \mathbf{H}_0  +  \eta  \ ( \pmb{\mathbf{\sigma}}_x k_y - \pmb{\mathbf{\sigma}}_y k_x) 
\end{equation}
where $\mathbf{H}_0$ represents the usual effective mass Hamiltonian, $\eta$ is the Rashba coefficient, $\pmb{\mathbf{\sigma}}$'s are Pauli spin matrices. We can think of the Rashba interaction  as an effective magnetic field that depends on the momentum of an electron if we associate $ | \eta \ \vec{k} | $ as the magnitude of the effective magnetic field. Indeed, injecting $x$ directed spins by using ferromagnetic contacts with  momentum ($\pm k_x$) can cause periodic oscillations of their spin, since these electrons would feel an effective magnetic field in the $y$ direction. This effect was initially proposed in Ref.~\cite{datta1990electronic} and experimentally confirmed in Ref.~\cite{nitta1997gate}, followed later by full demonstrations including the contacts \cite{koo2009control,wunderlich2010spin,choi2015electrical}.

Here, our purpose is to simply show how the NEGF formalism can be useful in modeling this effect in the presence of \textit{momentum, phase and spin} relaxations in a toy model.  Fig.~\ref{fig:rashba}a shows an NEGF simulation of 1D wire with Rashba spin orbit coupling, attached to two ferromagnetic leads that point in the $x$-direction.  The left lead is an $x$-directed ferromagnet that injects $x$-spins in the channel. In the presence of momentum and phase relaxation, the chemical potential for charge goes steadily down as in Fig.~\ref{fig:figure4}. We can calculate the corresponding \textit{spin} potentials by:
\[   f^{x,y,z}_{\ell}  = \frac{\text{Trace} \left[\mathbf{G}^n_{{\ell},{\ell}} \  \pmb{\mathbf{\sigma}}_{x,y,z}\right]}{\text{Trace} \left[\mathbf{A}_{{\ell},{\ell}}\right]}
\]
where `$\text{Trace}$' indicates tracing the $ 2\times2 $ blocks of $\mathbf{G}^n$ and $\mathbf{A}$ at lattice point $\ell$ and $\pmb{\mathbf{\sigma}}$'s are $2\times2$ Pauli spin matrices. Interestingly, if there is little to no spin relaxation, we see that the $x$-spins rotate in the channel even in the presence of high momentum and phase relaxation due to the effective magnetic field that is exerted on the $x$-spins  by the Rashba channel. In practice, however, momentum relaxation in high spin-orbit materials is known to give rise to  strong spin relaxation through the Elliott-Yafet \cite{steiauf2009elliott} mechanism. Our purpose is simply to show that it is possible to include spin relaxation \textit{independently} of momentum relaxation. This allows us to flexibly choose phase, momentum and spin relaxation rates to match experimentally measured quantities when modeling real experiments. As discussed in Ref.~\cite{datta2008nanoelectronic}, a possible spin flip relaxation model, $\pmb{\mathbf{\Sigma}}_{0}^{sf}$ can be written as: 
\begin{equation}
\pmb{\mathbf{\Sigma}}^{sf}_{0,i,j}=  D_0 (  \pmb{\mathbf{\sigma}}_x \mathbf{G}^n_{i,j} \pmb{\mathbf{\sigma}}_x   + \pmb{\mathbf{\sigma}}_y \mathbf{G}^n_{i,j} \pmb{\mathbf{\sigma}}_y  + \pmb{\mathbf{\sigma}}_z\mathbf{G}^n_{i,j} \pmb{\mathbf{\sigma}}_z)
\end{equation}
where the $\pmb{\mathbf{\Sigma}}_0^{sf}$ acts on all $2\times 2$ sub-blocks of the $\mathbf{G}^n$ matrix that are labeled $\mathbf{G}^n_{i,j}$. By multiplying the terms out, we can easily see the effect of this dephasing is to reinject an electron \textit{with an opposite spin} back to the channel that relaxes spin. In Fig.~\ref{fig:rashba}c we show how a pure spin relaxation model can be added in addition to momentum and phase breaking  to reduce the spin signals in the channel.

\subsection{Application to superconducting devices}
An important application of the NEGF method is to devices where one of the contacts or the channel is superconducting. This is of great current interest due to the excitement surrounding the search for Majorana bound states (MBS) or equivalently Majorana zero modes (MZM) in Rashba nanowire systems in proximity with superconductors\index{superconductor} \cite{Mourik1003,Lutchyn2010, Zhang2018, Zhang2019, Ren2019}.  It would take us too far afield to discuss the significant conceptual extensions needed for these problems and so we will refer interested readers to a couple of old papers \cite{Samanta1998,DATTA1999Can} and some recent ones \cite{San_Jose2013, Levy1995, Sriram2019 }.

\section{Frequently asked questions}
\label{sec:FAQ}

\subsection{Shouldn't we also consider the Poisson equation for modeling real devices?}

Yes. Earlier we mentioned the role of electron-electron interactions in causing dephasing which can be included through the anti-Hermitian part of the self-energy function $\pmb{\mathbf{\Sigma}}_{0}$. A far more important effect of electron-electron interactions is manifested through the Hermitian part of $\pmb{\mathbf{\Sigma}}_{0}$ which effectively modifies the Hamiltonian $\mathbf{H}$. The most important part of $\pmb{\mathbf{\Sigma}}_{0}$ is just the Coulomb potential $U$ arising from the electron density $\mathbf{G}^n/2 \pi$, and can be obtained by solving the Poisson equation and is often called the Hartree term. In addition there are more subtle contributions to  $\pmb{\mathbf{\Sigma}}_{0}$ generally referred to as exchange and correlation effects which subtract from the Hartree term, see for example, Chapter 2 of Ref.~\cite{datta2005quantum}.

Since $\pmb{\mathbf{\Sigma}}_{0}$ depends on $\mathbf{G}^n$, this requires a self-consistent calculation:  We start by assuming $\pmb{\mathbf{\Sigma}}_{0} = 0$, use Eq.~\ref{eq:eq1},\ref{eq:eq2} to calculate $\mathbf{G}^n$, use $\mathbf{G}^n$ to calculate  $\pmb{\mathbf{\Sigma}}_{0} = 0$, recalculate $\mathbf{G}^n$ from Eq.~\ref{eq:eq1},\ref{eq:eq2} and so on till the results converge. This requirement for an iterative solution makes it much more difficult to calculate the energy levels even for Helium which has two electrons compared to Hydrogen which has only one electron and hence no electron-electron interactions.

Two more points to note. Firstly, in device simulations it is common to use the Hamiltonian $\mathbf{H}$ from semi-empirical methods that already include the $\pmb{\mathbf{\Sigma}}_{0}$ for a uniform channel material in equilibrium. As such it is necessary to include \emph{only} the \emph{change} in $\pmb{\mathbf{\Sigma}}_{0}$ when the channel is driven out-of-equilibrium by the applied bias. Secondly, this correction is not important for low applied voltages, but plays a very significant role at high voltages. Indeed one can say that the saturation current for transistors is largely controlled by the Hartree term, making it primarily an electrostatically controlled device.

\subsection{Can you arbitrarily designate the ends as contacts?}

Designating a region as a contact\index{contact} implies that there is a significant increase in the number of conducting channels due to an increase in width and/or in the density of states\index{density of states}. When analyzing a given structure, it is important to scrutinize the results carefully, to make sure that the assumed contacts are indeed functioning as nearly ideal reservoirs.

\begin{figure}[!t]
\centering
	\includegraphics[width=0.6\linewidth,keepaspectratio]{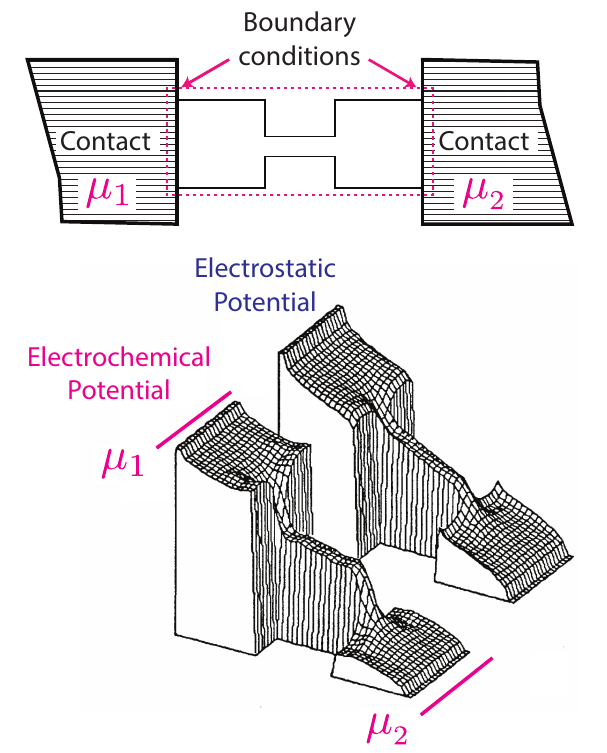}
	\caption{A wide-narrow-wide ballistic constriction where the contacts are imposed inside the widest region. A numerical simulation with phase scattering shows the 2D profile of electrochemical and electrostatic potentials. Reprinted Fig. 27b of Ref.~\cite{mclennan1991} with permission from Michael J. McLennan, Yong Lee, and Supriyo Datta
, Physical Review B, 43, 13846, 1991 Copyright (1991) by the American Physical Society.}
	\label{fig:figure6}
	\end{figure}

For example, Fig.~\ref{fig:figure6} is an excerpt from an old paper showing a wide-narrow-wide structure (see Fig.~27a of Ref.~\cite{mclennan1991}). We impose the contact boundary conditions\index{boundary conditions} inside the widest region as shown. Note the sharp drops in the occupation factor (or electrochemical potential)\index{electrochemical potential} at the narrow-wide interfaces due to the interface resistance\index{interface resistance}. The electrostatic potential\index{electrostatic potential} is obtained by convolving the electrochemical potential with a screening function as explained in the paper (Fig.~27b of Ref.~\cite{mclennan1991}). Note also the small drops where the wide regions meet the contacts indicating a small interface resistance.

One could save computation time by imposing the contact in the wide region rather than in the widest region. This may or may not be acceptable depending on how conductive the wide region is relative to the narrow region. A related phenomenon called the ``source-starvation''\index{source-starvation} effect where contact-like wide regions are driven out of equilibrium has been discussed in the context of ballisitic nanotransistors \cite{fischetti2007simulation}.

\subsection{Does NEGF give the correct coherent and semiclassical limits?}

Yes, with small dephasing\index{dephasing}, it matches the scattering theory for coherent transport, also called the Landauer-B\"{u}ttiker formalism\index{Landauer-B\"{u}ttiker formalism}. With sufficient dephasing, it reproduces results from the semiclassical Boltzmann method, thus providing a bridge from quantum to semiclassical transport. The example presented earlier shows how the NEGF result for coherent transport\index{coherent transport} includes oscillations around the Boltzmann result, which damp out when dephasing is included. Many such examples are included in Ref.~\cite{datta2012lessons} essentially as homework problems that are easily reproduced.

\subsection{ How is this related to the Kubo formalism?}
The Kubo formalism\index{Kubo formalism} is a linear response theory based on small perturbations close to equilibrium. Many of the problems in modern nanoelectronics require a framework that can handle transport far from equilibrium. The results from NEGF agree with the Kubo formalism in the linear response regime close to equilibrium.

\subsubsection{Isn't it a problem to have separate electrochemical potentials in a single system?}
On the contrary, we would argue that current flow requires separate electrochemical potentials\index{electrochemical potential}, just as the flow of heat requires separate temperatures. There is nothing fundamentally wrong with invoking two large reservoirs held at two different electrochemical potentials and/or temperatures connected by a nanoscale conductor.

\subsubsection{Doesn't the Kubo formula use a single electrochemical potential?}
Yes, the Kubo formalism calculates an equilibrium quantity, namely the noise\index{noise}, and relates it to the linear transport coefficients through the fluctuation-dissipation theorem\index{fluctuation-dissipation theorem}. But this only works for transport close to equilibrium and is not possible in general. NEGF on the other hand addresses the non-equilibrium problem directly. Like the Boltzmann equation, NEGF too permits the use of separate electrochemical potentials as boundary conditions.

\subsubsection{Can't you have equilibrium currents with a single electrochemical potential?}
Yes, but these are distinct from the transport currents we are discussing. For example in the quantum Hall regime\index{quantum Hall regime}, there are circulating edge currents\index{edge currents} which involve all electrons and not just the ones  with energies close to $\mu$. But these are normally not measured with external probes.

\subsection{Isn't the flow of electricity essentially a many-body process?}

\begin{figure}[!ht]
\sidecaption
\includegraphics[width=0.6\linewidth,keepaspectratio]{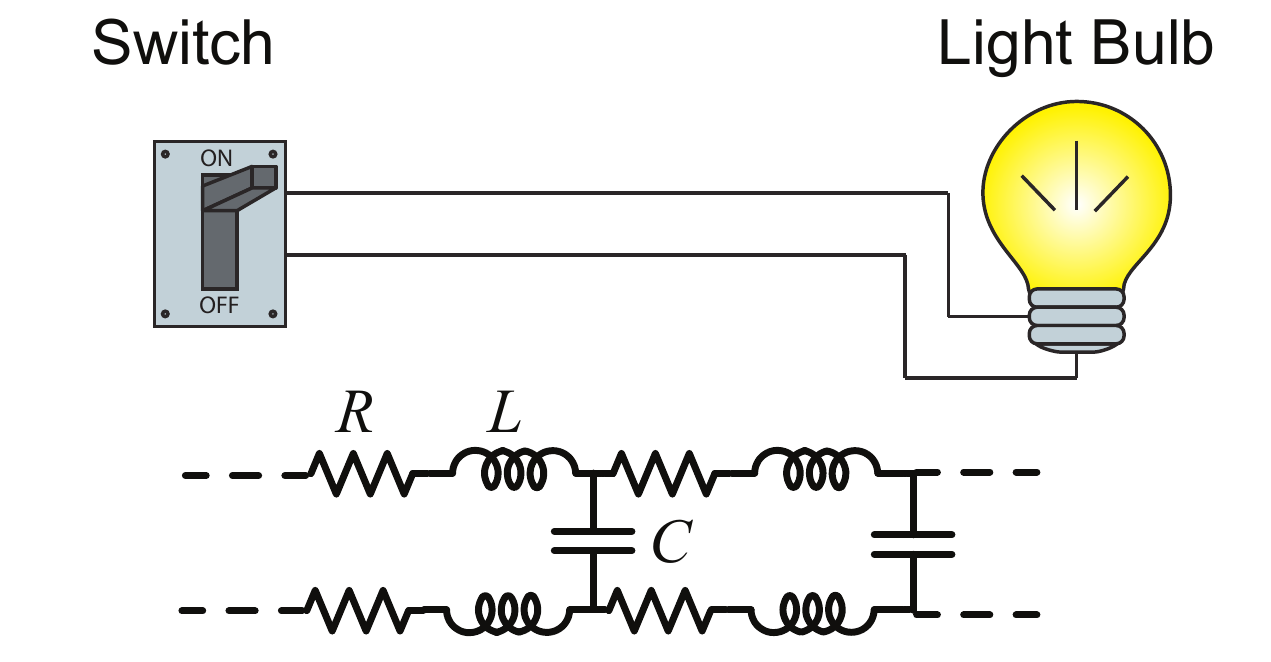}\\
\caption{ When we turn the switch on a signal propagates to the light bulb which can be described by a  distributed $RLC$ model. The capacitance $C$ and the inductance $L$ are determined primarily by electrostatics and magnetostatics respectively.}
\label{signal}
\end{figure}

As undergraduates we learn that current flow, at least in metals, is a many-body interacting process: When we turn the switch on, a signal propagates to the light bulb at the speed of light. One electron pushes the next electron, which pushes the next electron and so on. If we had to wait for an electron to transmit from the switch to the bulb it would take far too long. And so it might seem futile to describe current flow without invoking many-body interactions. However, we believe that while signal propagation cannot be analyzed without interactions, the low frequency resistance can be.

This signal propagation can be described quantitatively by a distributed $RLC$ model \cite{salahuddin2005transport}\begin{footnote}{For a spin generalization of similar ideas, see \cite{sayed2018transmission}.} \end{footnote}. The $L$'s and $C$'s in this model have a transport component, but usually represent primarily the interactions and the velocity is given by $1/\sqrt{LC}$. However, these parameters are not involved in determining the steady-state or  $dc$ current since at low frequencies the $L$'s appear as short circuits, and the $C$'s are open circuits. The low frequency resistance $R$ can be understood at least approximately without invoking interactions.

This observation appears to apply to quantum transport as well. For example, in 1985 Richard Webb and collaborators from IBM reported \cite{webb1985observation} the Aharonov-Bohm (AB) effect\index{Aharonov-Bohm effect} in metallic rings with a period described quite well by the one-electron theory without interactions, as if we were talking about AB paths in vacuum! The one-electron picture works so well possibly because there is a Landau-style (We mean Landau, not Landauer!) dressed quasi-particle that ``actually'' moves from left to right. We have some mental pictures of what such a quasi-electron looks like, but we are not aware of any serious discussions along these lines.

\subsubsection{Are many-body effects irrelevant in transport experiments?}

That is not at all what we mean. Our point is that many-body effects are fairly obvious at high frequencies starting with the well-known fact that signals travel at the speed of light and not at the speed of electrons. By contrast their role at low frequencies are more subtle, like Coulomb blockade and the Kondo effect. More recently experiments on graphene have revealed intriguing evidence of collective motion \cite{kumar2017superballistic}.

In a volume published on the occasion of the 50th anniversary of $Anderson$ $localization$, Phil Anderson commented ``\ldots What might be of modern interest is the channel concept which is so important in localization theory. The transport properties at low frequencies can be reduced to a sum over one dimensional channels.''

Note the qualification \emph{low frequencies}. Transmission of an electron from the left to the right contact is essential only for steady-state charge transport. High frequency currents can flow capacitively without end-to-end transfer of electrons. Also, experiments have revealed that spins can be transported across insulators by magnons \index{magnon} without the actual transmission of electrons\begin{footnote}{See for example Ref. \cite{kajiwara2010transmission} and \cite{sayed2016spin} for possible practical applications of such pure spin conductors.}\end{footnote}.

\subsubsection{Would NEGF be suitable for describing such effects?}
NEGF provides a clear prescription for including arbitrary interactions to any desired order in perturbation theory. But some effects may require us to go beyond MBPT: perhaps a non-perturbative approach, or a different perturbation parameter. Single-electron charging effects\index{single electron charging effect} provide a good example \cite{Natalya2008}.

\subsection{How can we teach NEGF without advanced quantum statistical mechanics?}

The NEGF eqs.~(\ref{eq:eq1}-\ref{eq:eq2}) are essentially the same as Eqs.~(75)-(77) of L. V. Keldysh, Sov. Phys. JETP, vol. 20, p. 1018 (1965) \cite{keldysh1965diagram}, which is one of the seminal  papers on the NEGF method that obtained these equations using MBPT and established the ``diagram technique for non-equilibrium processes'' for calculating the self-energy functions $\mathbf{\Sigma}_{0}$ and $\mathbf{\Sigma}^{\text{in}}_{0}$. Much of the literature is based on the original MBPT-based approach and this makes it inaccessible to those unfamiliar with advanced quantum statistical mechanics.

We obtain the NEGF equations directly from a one-electron Schr\"{o}dinger equation\index{one-electron Schr\"{o}dinger equation} using relatively elementary arguments. These equations have been used to discuss many problems of great interest like quantized conductance, (integer) quantum Hall effect, Anderson localization\index{Anderson localization}, resonant tunneling and spin transport without a systematic treatment of many-body effects. But it goes beyond purely coherent transport allowing us to include phase-breaking interactions (both momentum-relaxing and momentum-conserving) within a self-consistent Born approximation\index{Born Approximation}.

The NEGF equations provide a unified framework for such problems, spanning a wide range of materials and phenomena all the way from molecular to ballistic to diffusive transport and has been widely adopted by the nanoelectronics community for device analysis and design.

\vspace{-10pt}
\subsection{But is this the real NEGF?}
\vspace{-10pt}

\begin{itemize}
\item The answer is NO, if we associate NEGF with the MBPT commonly used to obtain $\pmb{\mathbf{\Sigma}}_{0}$ and $\pmb{\mathbf{\Sigma}}^{\text{in}}_{0}$ appearing in Eqs.~(\ref{eq:eq1}-\ref{eq:eq3}) 

\item The answer is YES, if we associate NEGF with Eqs.~(\ref{eq:eq1}-\ref{eq:eq3}) irrespective of how the $\pmb{\mathbf{\Sigma}}$'s are obtained.
\end{itemize}

\noindent Which answer we choose is clearly a matter of perspective, but the second viewpoint seems more in keeping with semiclassical transport theory, where the (steady-state) Boltzmann approach is identified with the equation:
\[\pmb{\mathbf{\nu }}\,\cdot\,{\nabla }f+\mathbf{F}\,\cdot\,{\nabla }_{p} f=\; S_{\text{op}} f \]
\noindent and NOT with the evaluation of the scattering operator\index{scattering operator} $S_{\text{op}}$ which is analogous to the $\pmb{\mathbf{\Sigma}}$'s in NEGF. While the concept of the collision integral involved in $S_{\text{op}}$ was well-known, the quantitative details of its evaluation has evolved significantly since the days of Boltzmann. Similarly, totally new approaches that go beyond many-body perturbation theory (MBPT) for evaluating the $\pmb{\mathbf{\Sigma}}$'s have been and will be developed as we apply NEGF to newer problems.
\vspace{-15pt}

\subsection{Contact-ing Schr\"{o}dinger}
\vspace{-10pt}

In summary, we feel that the scope and utility of the NEGF equations Eq.~(\ref{eq:eq1}-\ref{eq:eq3})  transcend the MBPT-based approach originally used to derive it. NEGF teaches us how to combine quantum dynamics with ``contacts'' much as Boltzmann taught us how to combine classical dynamics with ``contacts'', using the word ``contacts'' in a broad figurative sense to denote all kinds of entropy-driven processes. There are many problems of fundamental and applied significance that can be simulated straightforwardly within the NEGF method without a first principles treatment of many-body effects which in any case would be perturbative in nature. This includes novel phenomena like the (integer) quantum Hall effect as well as new devices like magnetic tunnel junctions. As such, we feel that this elegant framework for ``contact-ing'' the Schr\"{o}dinger  equation should be of broad interest to anyone working on device physics or non-equilibrium statistical mechanics in general. \\

\footnotesize
\begin{programcode}{Python codes for Figures~\ref{fig:figure3} and \ref{fig:figure4}}
\begin{lstlisting}[language=iPython]
# This code illustrates how to perform NEGF calculations on a simple	
# 1D device with one scatterer in the middle of the channel region.
import numpy as np
import matplotlib.pyplot as plt
fontPlan = {'family': 'sans','weight': 'bold','size': ¢12¢,'color':'darkred',}
	
t0 = ¢1¢; Np = ¢51¢; X = np.arange(¢0¢,Np,¢1¢); Nh = int(Np/¢2¢); zplus = ¢1e-12¢*¢1j¢
L = np.diag(np.append(¢1¢, np.zeros(Np-¢1¢)))
R = np.diag(np.append(np.zeros(Np-¢1¢), ¢1¢))                 
option = 'coherent' # This generates the bottom plot in Fig. 3
# use option = 'phase-relaxation' for Fig. 5 (bottom left) 
#or option = 'both-relaxation' for Fig. 5 (bottom right)
	
DList = np.array([¢0¢, ¢9e-2¢*t0**¢2¢]); sigB = np.zeros(Np); siginB = np.zeros(Np)
H0 = ¢2¢*t0*np.diag(np.ones(Np))- t0*np.diag(np.ones(Np -¢1¢),¢1¢) \
- t0*np.diag(np.ones(Np-¢1¢),-¢1¢)          # 1D Hamiltonian
N1 = Nh; UB1 = ¢1¢*t0; H0[N1][N1] = H0[N1][N1] + UB1; H = H0
EE = t0; ck = (¢1¢-(EE+zplus)/(¢2¢*t0)); ka = np.arccos(ck); v = ¢2¢*t0*np.sin(ka)

# Semiclassical profile
T = np.real(v**¢2¢/(UB1**¢2¢+v**¢2¢)); R1 = (¢1¢-T)/T   
D = DList[¢0¢] if option.lower() != 'both-relaxation' else DList[¢1¢]	
TT = np.real(v**¢2¢/(D+v**¢2¢)); R2 = ¢1¢*(¢1¢-TT)/TT 
RR = np.append(¢0.5¢,np.append(R2*np.ones(Nh),R1))
RR = np.append(RR,np.append(R2*np.ones(Nh),¢0.5¢))
RR = np.cumsum(RR); Vx = np.ones(Np+¢2¢) -(RR/RR[Np+¢1¢]); Fclass = Vx[¢1¢:Np+¢1¢]
	
# Based on resistance estimates
s1 = -t0*np.exp(¢1j¢*ka); sig1 = np.kron(L,s1); ck =(¢1¢-(EE+zplus)/(¢2¢*t0))
ka = np.arccos(ck); s2 = -t0*np.exp(¢1j¢*ka); sig2 = np.kron(R,s2)
gam1 = ¢1j¢*(sig1-np.conj(sig1.T)); gam2 = ¢1j¢*(sig2-np.conj(sig2.T))	
G = np.linalg.inv((EE*np.eye(Np))-H-sig1- sig2)
Tcoh = np.real(np.trace(gam1*G*gam2*np.conj(G.T)))
	
if option.lower()!='phase-relaxation': ff = lambda x,y: np.diag(np.diag(x*y))
else: ff = lambda x,y: x*y;
D = DList[¢0¢] if option.lower()=='coherent' else DList[¢1¢]
change = ¢100¢
while change > ¢1e-6¢:
  G = np.linalg.inv((EE*np.eye(Np)) - H - sig1 - sig2 - sigB)
  sigBnew = ff(D,G); change = np.sum(np.sum(np.abs(sigBnew - sigB)))
  sigB = sigB + ¢0.25¢*(sigBnew - sigB)	
A = np.real(np.diag(1j*(G-np.conj(G.T)))); change = ¢100¢
while change > ¢1e-6¢:
  Gn = np.matmul(G,np.matmul((gam1 + siginB),np.conj(G.T)))
  siginBnew = ff(D,Gn); change = np.sum(np.sum(np.abs(siginBnew - siginB)))
  siginB = siginB + ¢0.25¢*(siginBnew - siginB)
F = np.real(np.diag(Gn))/A

Xclass = np.append(np.arange(-¢10¢,¢0¢,¢1¢), np.append(-¢0.001¢,X))
Xclass = np.append(Xclass, np.append(¢50.001¢,np.arange(¢51¢,¢61¢,¢1¢)))
FclassFull = np.append(np.ones(¢11¢), np.append(Fclass,np.zeros(¢11¢)))

plt.plot(X,F,'k-', label='NEGF', linewidth=¢1.0¢)
plt.plot(Xclass,FclassFull,'r-', linewidth=¢1.5¢)
plt.plot(X,Fclass,'ro-', label='Semiclassical', linewidth=¢1.0¢,markersize=¢2¢)
plt.xticks(fontsize=¢10¢); plt.yticks(fontsize=¢10¢); plt.xlim(-¢10¢,¢60¢)
plt.ylim(-¢0.01¢,¢1.01¢); plt.xlabel('z',fontdict=fontPlan)
plt.ylabel('f',fontdict=fontPlan)
plt.legend(frameon=False,fontsize = ¢10¢, loc = 'best'); plt.show()
\end{lstlisting}
\end{programcode}
\begin{programcode}{Python code for Fig.~\ref{fig:rashba}}
\begin{lstlisting}[language=iPython]
# This code illustrates how to perform NEGF calculations on a simple 1D device 
# with Rashba channel.
import numpy as np
import matplotlib.pyplot as plt
fontPlan = {'family': 'sans','weight': 'bold','size': ¢12¢,'color':'darkred',}

LL = ¢200e-10¢; h = ¢6.62e-34¢/¢2¢/np.pi; a = ¢1e-10¢; NM = int(np.ceil(LL/a));
m = ¢9e-31¢; q = ¢1.6e-19¢; t0 = h**¢2¢/(¢2¢*m*a**¢2¢)/q; I2 = np.eye(¢2¢);
Pauli_y = np.array([[¢0¢+¢0j¢, ¢0¢-¢1j¢], [¢0¢+¢1j¢, ¢0¢+¢0j¢]])
Pauli_z = np.array([[¢1¢+¢0j¢, ¢0¢+¢0j¢], [¢0¢+¢0j¢, ¢-1¢+¢0j¢]])
Pauli_x = np.array([[¢0¢+¢0j¢, ¢1¢+¢0j¢], [¢1¢+¢0j¢, ¢0¢+¢0j¢]])

SZ = np.kron(np.eye(NM),Pauli_z); SX = np.kron(np.eye(NM), Pauli_x); 
SY = np.kron(np.eye(NM), Pauli_y); SI = np.kron(np.eye(NM), I2);
S0 = np.mod(np.kron(np.eye(NM), np.array([[¢1¢, ¢1¢], [¢1¢, ¢1¢]])) + \ 
np.ones((¢2¢*NM, ¢2¢*NM)), ¢2¢);
S1 = np.kron(np.eye(NM), np.array([[¢1¢, ¢1¢], [¢1¢, ¢1¢]])); 
S11 = np.ones((¢2¢*NM, ¢2¢*NM));

DP = ¢0.09¢*t0**¢2¢ ;DM = ¢0.15¢*¢t0¢**¢2¢; DS = ¢0.00125¢/¢4¢*t0**¢2¢; eta = ¢0.2¢*t0;
alpha = np.array([[¢2¢*t0, ¢0¢], [¢0¢, ¢2¢*t0]]);
gamma = np.array([[-t0, ¢0¢], [¢0¢, -t0]]) - eta*Pauli_y/¢2j¢

H = np.kron(np.eye(NM), alpha) + np.kron(np.diag(np.ones(NM - ¢1¢), ¢-1¢),gamma) \
+ np.kron(np.diag(np.ones(NM - ¢1¢), +¢1¢),np.conj(gamma.T))

zplus = ¢1j¢*¢1e-12¢; EE = t0; ka = np.arccos(¢1¢ - (EE + zplus)/(¢2¢*t0));
px1 = ¢0.95¢; py1 = ¢0.0¢; pz1 = ¢0.0¢; px2 = ¢0.95¢; py2 = ¢0.0¢; pz2 = ¢0.0¢;
sigma1 = -t0*np.exp(¢1j¢*ka)*(I2 + pz1*Pauli_z + px1*Pauli_x + py1*Pauli_y);
sigma2 = -t0*np.exp(¢1j¢*ka)*(I2 + pz2*Pauli_z + px2*Pauli_x + py2*Pauli_y);

LL = np.zeros(NM); RR = np.zeros(NM); LL[¢0¢] = ¢1¢; RR[NM-¢1¢] = ¢1¢;
Sigma1 = np.kron(np.diag(LL), sigma1); Sigma2 = np.kron(np.diag(RR), sigma2);
SigmaB = ¢0¢*Sigma1; SigmaIn = ¢0¢*Sigma1; 
Gamma1 = ¢1j¢*(Sigma1 - np.conj(Sigma1.T)); 
Gamma2 = ¢1j¢*(Sigma2 - np.conj(Sigma2.T));

f1 = ¢1¢; f2 = ¢0¢; error = ¢1¢;
while error > ¢1e-5¢: 
  GR = np.linalg.inv(EE*np.eye(¢2¢*NM) - H - Sigma1 - Sigma2 - SigmaB);
  SigmaBnew = DS*(SZ@(GR)@SZ + SX@(GR)@SX + SY@(GR)@SY) + DM*(S1*GR) \
  + DP*(S11*GR);
  error = np.sum(np.abs(SigmaBnew - SigmaB)) \ 
  /np.sum(np.abs(SigmaBnew + SigmaB + ¢1e-15¢));
  SigmaB = SigmaB + ¢0.85¢*(SigmaBnew - SigmaB);
GA = np.conj(GR.T); AA = 1j*(GR - GA);

error = ¢1¢;
while error > ¢1e-6¢:
  Gn = GR@(Gamma1*f1 + Gamma2*f2 + SigmaIn)@GA;
  SigmaInNew = DS*(SZ@(Gn)@SZ + SX@(Gn)@SX + SY@(Gn)@SY) + DM*(S1*Gn) \ 
  + DP*(S11*Gn);
  error = np.sum(np.abs(SigmaInNew - SigmaIn)) \ 
  /np.sum(np.abs(SigmaInNew + SigmaIn + ¢1e-15¢));
  SigmaIn = SigmaIn + ¢0.85¢*(SigmaInNew - SigmaIn);

 
denom = np.sum(np.reshape(np.diag(AA@SI),(NM,¢2¢)),axis = ¢1¢);
charge = np.real(np.sum(np.reshape(np.diag(SI@Gn),(NM,¢2¢)),axis = ¢1¢)/denom)
spin_x = np.real(np.sum(np.reshape(np.diag(SX@Gn),(NM,¢2¢)),axis = ¢1¢)/denom)
spin_y = np.real(np.sum(np.reshape(np.diag(SY@Gn),(NM,¢2¢)),axis = ¢1¢)/denom)
spin_z = np.real(np.sum(np.reshape(np.diag(SZ@Gn),(NM,¢2¢)),axis = ¢1¢)/denom)


coef = np.polyfit(np.arange(¢0¢,NM,¢1¢),charge,¢1¢); poly1d_fn = np.poly1d(coef);
X = np.arange(¢0¢,NM,¢1¢); Fclass = poly1d_fn(X);
XFull = np.append(np.arange(¢-10¢,¢0¢,¢1¢), np.append(¢-0.001¢,X))
XFull = np.append(XFull , np.append(¢200.001¢,np.arange(¢201¢,¢211¢,¢1¢)))
FclassFull = np.append(np.ones(¢11¢), np.append(Fclass ,np.zeros(¢11¢)))

plt.plot(X,charge, label='NEGF: charge', linewidth=¢1.5¢, alpha=¢0.9¢, color=(¢0.92157¢,¢0.384366¢,¢0.207899¢))
plt.plot(XFull, FclassFull, '--k',label='semiclassical',linewidth=¢1.5¢)
plt.plot(X,spin_z, label='NEGF: z-spin', linewidth=¢1.5¢, color=(¢0.560784¢,¢0.690196¢,¢0.196218¢))
plt.plot(X,spin_x, label='NEGF: x-spin', linewidth=¢1.5¢, color=(¢0.368627¢,¢0.505882¢,¢0.709803¢))
plt.plot(X,spin_y, label='NEGF: y-spin', linewidth=¢1.5¢, color=(¢1¢,¢0¢,¢0¢))
plt.xticks(fontsize=¢10¢); plt.yticks(fontsize=¢10¢); plt.xlim(¢-20¢,¢240¢)
plt.ylim(¢-0.45¢,¢1.1¢); plt.xlabel('x',fontdict=fontPlan);
plt.ylabel('f',fontdict=fontPlan)
plt.legend(frameon=False,fontsize = ¢8¢, loc = 'best'); plt.show()
\end{lstlisting}
\end{programcode}

\section*{Acknowledgments}
The authors thank Bhaskaran Muralidharan for valuable discussions related to the applications of NEGF to modern superconducting devices.


\begingroup
\let\cleardoublepage\clearpage
\titleformat{\chapter}[runin]{}{}{}{}
\titlespacing{\chapter}
{0pt}{-\baselineskip}{0pt}
\endgroup

\end{document}